\documentclass[a4paper,11pt]{article}
\pdfoutput=1

\usepackage{jheppub}
\usepackage[T1]{fontenc}
\usepackage[italic]{hepparticles}
\usepackage{hepunits}
\usepackage{hepnames}
\usepackage{soul}
\usepackage[dvipsnames]{xcolor}

\def\bTosmumu{\ensuremath{\Pbeauty \to \Pstrange \APmuon \Pmuon}\xspace}
\def\bTosgamma{\ensuremath{\Pbeauty \to \Pstrange \Pphoton}\xspace}
\def\bTosll{\ensuremath{\Pbeauty \to \Pstrange \APlepton \Plepton}\xspace}

\def\bTos{\ensuremath{\Pbeauty \to \Pstrange }\xspace}

\def\l{\ensuremath{\mathcal{L}}}
\def\cov{\ensuremath{{\rm Cov}}}
\def\BtoKstarGamma{\ensuremath{{ \HepParticle{B}{}{} \to \PKstarzero \gamma}}}
\def\PKstarzero{\HepParticle{K}{}{*0}}
\def\PKstarplus{\HepParticle{K}{}{*+}}
\def\BtoKstaree{\ensuremath{\HepParticle{B}{d}{} \to \PKstarzero \APelectron \Pelectron}\xspace}

\def\BdToKstarmumu{\ensuremath{\HepParticle{B}{d}{} \to \PKstarzero \APmuon \Pmuon}\xspace}
\def\BuToKstarmumu{\ensuremath{\HepParticle{B}{u}{} \to \PKstarplus \APmuon \Pmuon}\xspace}

\def\BdToKstaree{\ensuremath{\HepParticle{B}{d}{} \to \PKstarzero \APelectron \Pelectron}\xspace}
\def\BToXsmumu{\ensuremath{\PB \to X_s \APmuon \Pmuon}\xspace}
\def\BTomumu{\ensuremath{\HepParticle{B}{s}{} / \HepParticle{B}{d}{} \to \APmuon \Pmuon}\xspace}
\def\BsTomumu{\ensuremath{\HepParticle{B}{s}{} \to \APmuon \Pmuon}\xspace}
\def\BdTomumu{\ensuremath{\HepParticle{B}{d}{} \to \APmuon \Pmuon}\xspace}

\def\BsTophimumu{\ensuremath{\HepParticle{B}{s}{} \to \Pphi \APmuon \Pmuon}\xspace}
\def\BToKmumu{\ensuremath{\PB \to \PK \APmuon \Pmuon}\xspace}
\def\BToKee{\ensuremath{\PB \to \PK \APelectron \Pelectron}\xspace}

\def\LbToLmumu{\ensuremath{\PLambdab \to \PLambda \APmuon \Pmuon}\xspace}

\def\rk{\ensuremath{R_{\PK}}\xspace}
\def\rkstar{\ensuremath{R_{\PKstar}}\xspace}

\renewcommand{\Re}{\operatorname{Re}}

\def\recSeven{\ensuremath{\Re(C_{7})}\xspace}
\def\recNine{\ensuremath{\Re(C_{9})}\xspace}
\def\recTen{\ensuremath{\Re(C_{10})}\xspace}
\def\recDSeven{\ensuremath{\Re(\Delta C_{7})}\xspace}
\def\recDNine{\ensuremath{\Re(\Delta C_{9})}\xspace}
\def\recDTen{\ensuremath{\Re(\Delta C_{10})}\xspace}

\def\p5prime{\ensuremath{P^{\prime}_5}\xspace}
\def\qsq{\ensuremath{q^2}\xspace}
\def\gevgevcccc{\ensuremath{\rm GeV^2/c^4}\xspace}

\newcommand\tabref[1]{Tab.~\ref{#1}}
\newcommand\figref[1]{Fig.~\ref{#1}}
\newcommand{\secref}[1]{Sec.~\ref{#1}\xspace}

\newcommand{\av}[1]{\langle #1 \rangle}
\newcommand{\bin}{{\rm bin}}
\newcommand{\WC}{Wilson coefficient\xspace}
\newcommand{\WCs}{Wilson coefficients\xspace}
\def\sm{SM\xspace}

\newcommand{\gambit}{\textsf{GAMBIT}\xspace}
\newcommand{\flavbit}{\textsf{FlavBit}\xspace}
\newcommand{\scannerbit}{\textsf{ScannerBit}\xspace}

\newcommand\superisofour{\textsf{SuperIso 4.1}\xspace}
\newcommand\heplike{\textsf{HEPLike}\xspace}
\newcommand\heplikedata{\textsf{HEPLikeData}\xspace}
\newcommand\diver{\textsf{Diver}\xspace}
\newcommand\twalk{\textsf{T-Walk}\xspace}
\newcommand\pippi{\textsf{pippi}\xspace}

\allowdisplaybreaks

\begin{document}
\def\thefootnote{\fnsymbol{footnote}}

\begin{center}

\vspace*{1.cm}{\Large\bf {A model-independent analysis of $\bTosmumu$ transitions\\[0.2cm] with \gambit's \flavbit}}

\setlength{\textwidth}{11cm}

\vspace{2.cm}
{\large\bf
Jihyun Bhom$^{a}$
,\,
Marcin Chrzaszcz$^{a}$
,\,
Farvah~Mahmoudi$^{b,c}$
,\,
Markus T.~Prim$^{d}$
,\,
Pat Scott$^{e,f}$
,\,
Martin White$^{g}$
}

\vspace{1.cm}
{\em $^a$Henryk Niewodniczanski Institute of Nuclear Physics Polish Academy of Sciences, Krakow, Poland}\\[0.2cm]
{\em $^b$Univ Lyon, Univ Lyon 1, CNRS/IN2P3, Institut de Physique Nucl\'eaire de Lyon,\\ UMR5822, F-69622 Villeurbanne, France}\\[0.2cm]
{\em $^c$Theoretical Physics Department, CERN, CH-1211 Geneva 23, Switzerland}\\[0.2cm]
{\em $^d$Institute of Experimental Particle Physics, Karlsruhe Institute of Technology,\\ Karlsruhe, Germany}\\[0.2cm]
{\em $^e$School of Mathematics and Physics, The University of Queensland, St.\ Lucia,\\ Brisbane, QLD 4072, Australia}\\[0.2cm]
{\em $^f$Department of Physics, Imperial College London, SW7 2AZ, South Kensington, UK}\\[0.2cm]
{\em $^g$Department of Physics, University of Adelaide, Adelaide, SA 5005, Australia}\\[0.2cm]

\end{center}

\renewcommand{\thefootnote}{\arabic{footnote}}
\setcounter{footnote}{0}

\vspace{1.cm}
\thispagestyle{empty}
\centerline{\bf ABSTRACT}
\vspace{0.5cm}
The search for flavour-changing neutral current effects in $B$-meson decays is a powerful probe of physics beyond the Standard Model. Deviations from SM behaviour are often quantified by extracting the preferred values of the Wilson coefficients of an operator product expansion. We use the \flavbit module of the \gambit package to perform a simultaneous global fit of the Wilson coefficients $C_7$, $C_9$, and $C_{10}$ using a combination of all current data on \bTosmumu transitions. We further extend previous analyses by accounting for the correlated theoretical uncertainties at each point in the Wilson coefficient parameter space, rather than deriving the uncertainties from a Standard Model calculation. We find that the best fit deviates from the \sm value with a significance of 6.6$\sigma$. The largest deviation is associated with a vector coupling of muons to $b$ and $s$ quarks.

\newpage


\section{Introduction}
\label{sec:intro}

In the Standard Model (\sm), flavour-changing neutral currents (FCNC) are heavily suppressed by the Glashow-Iliopoulos-Maiani (GIM; \cite{PhysRevD.2.1285}) mechanism. Since the start of the LHC, experiments have observed numerous deviations from the predictions of the \sm in \bTosmumu transitions, starting with the 2013 LHCb collaboration observation of a deviation in the \p5prime observable in the range $\qsq \in \left[ 4.30, 8.68 \right]~\gevgevcccc$ of the decay $\HepParticle{B}{d}{} \to \PKstarzero \APmuon \Pmuon$~\cite{Aaij:2013qta}. Further discrepancies were later observed in measurements of the \BsTophimumu~\cite{Aaij:2013aln,LHCb:2021zwz}, \BToKmumu~\cite{Aaij:2016cbx}, and \LbToLmumu~\cite{Aaij:2015xza} decays in both the angular and branching fraction observables. Given the consistency of the observations, other experiments~\cite{Abdesselam:2016llu,Aaboud:2018krd,Sirunyan:2017dhj} have performed measurements of the \BdToKstarmumu decay, finding results consistent with the discrepancies seen earlier.

As has been studied in previous analyses, the discrepancies can be solved by reducing the $C_9$ \WC by one quarter of the \sm value (see for example Refs.~\cite{DAmico:2017mtc,Geng:2017svp,Hurth:2017hxg,Alguero:2019ptt,Alok:2019ufo,Ciuchini:2019usw,Aebischer:2019mlg,Kowalska:2019ley,Arbey:2018ics,Datta:2019zca,Arbey:2019duh}). Unfortunately, these processes suffer from non-factorisable corrections, making the size of the theoretical uncertainties, and therefore the overall significance of the results, difficult to quantify (see Refs.~\cite{Khodjamirian:2010vf,Khodjamirian:2012rm,Dimou:2012un,Lyon:2013gba,Lyon:2014hpa,Bobeth:2017vxj,Blake:2017fyh}, and \cite{Hurth:2020rzx} for a recent discussion of the hadronic corrections).

The \sm predicts that the rate of \bTosll transitions is independent of the flavour of the leptons involved, except for mass effects which are negligible when studying the first two generations $\Plepton = \Pe, \Pmu$.  In addition to the discrepancies seen purely with muons, the LHCb collaboration has therefore also performed explicit tests of lepton universality in \bTosll transitions. The first test is to measure the ratio $\rk\equiv\mathcal{B}(\BToKmumu)/\mathcal{B}(\BToKee)$, whilst a second is to measure $\rkstar = \mathcal{B}(\BdToKstarmumu)/\mathcal{B}(\BdToKstaree)$. In both cases the \sm prediction is known at the $1\%$ level, as the hadronic uncertainties cancel in taking the ratio of the two branching ratios. The LHCb experiment measured a lower value than the \sm prediction with a significance of $3.1\sigma$ and $2.1-2.5\sigma$ for \rk and \rkstar respectively~\cite{LHCb:2021trn,Aaij:2017vbb}.

In this paper, we focus mainly on tests with muons, exploring the extent to which the combination of all such data to date either constrain flavour-universal new physics (i.e.\ coupling identically to all leptons) in \bTos transitions, or prefer it in comparison to the \sm.  Using the \flavbit \cite{flavbit} module of the Global and Modular Beyond-Standard Model Inference Tool (\gambit; \cite{gambit,grev}), we carry out a model-independent analysis by simultaneously fitting three \WCs in an effective field theory for interactions of $b$ and $s$ quarks with leptons and photons.  We significantly improve on previous analyses by explicitly re-computing theory uncertainties at every \WC combination, rather than assuming that they are constant and given by their \sm values across the entire parameter space.  The final result is a 6.6$\sigma$ preference for new physics, overwhelmingly associated with the vector coupling of muons to $b$ and $s$ quarks.

We begin in \secref{sec:theory} with a description of the effective field theory framework in which we work, followed by explanations of the observables (\secref{sec:exp}) and likelihoods (\secref{sec:stats}) involved in our fits.  We show the results of our analysis in \secref{sec:results}, including preferred regions of the \WC parameter space and the spectrum of observables at our best-fit point, before concluding in \secref{sec:conclusions}.

\section{Theoretical Framework}
\label{sec:theory}

Our analysis is based on the effective Hamiltonian approach, in which the Operator Product Expansion is used to separate physics at low energies from a (possibly unknown) high energy theory. In this framework, the transition from an initial state $i$ to a final state $f$ is proportional to the squared matrix element $|\langle f |{\cal H}_{\rm eff}|i\rangle|^2$, with the effective Hamiltonian for $b \rightarrow  s$ transitions given by
\begin{equation}
\mathcal{H}_{\rm eff}  =  -\frac{4G_{F}}{\sqrt{2}} V_{tb} V_{ts}^{*} \sum_{i=1}^{10} \Bigl(C_{i}(\mu) \mathcal{O}_i(\mu)+C'_{i}(\mu) \mathcal{O}'_i(\mu)\Bigr)\;.
\end{equation}
$G_F$, $V_{tb}$ and $V_{ts}$ are \sm parameters (the Fermi constant and two CKM matrix elements, respectively), $\mu$ is the energy scale at which the calculation is being performed, and the $O_i$ are local operators providing low-energy descriptions of high-energy physics that has been integrated out. The operators each come with an associated \WC $C_i$ which, for a particular high-energy physics model, is calculable within the framework of perturbation theory.  This is done by matching the high-scale theory to the low-energy effective theory at a scale $\mu_W$, which is of the order of the $W$ boson mass. The renormalisation group equations of the low-energy effective theory can then be used to evolve the \WCs to the scale $\mu_b$, which characterises $B$ meson decay calculations and is thus of order $m_b$.

The operators that are most relevant for rare B decays featuring FCNCs are
\begin{align}
\label{physical_basis}
\mathcal{O}_1& =  (\bar{s} \gamma_{\mu} T^a P_L c) (\bar{c} \gamma^{\mu} T^a P_L b)\;,\nonumber\\
\mathcal{O}_2& = (\bar{s} \gamma_{\mu} P_L c) (\bar{c} \gamma^{\mu} P_L b)\;,  \nonumber\\
\mathcal{O}_3& =  (\bar{s} \gamma_{\mu} P_L b) \sum_q (\bar{q} \gamma^{\mu} q)\;, \nonumber\\
\mathcal{O}_4& = (\bar{s} \gamma_{\mu} T^a P_L b) \sum_q (\bar{q} \gamma^{\mu} T^a q)\;,\nonumber\\
\mathcal{O}_5& =  (\bar{s} \gamma_{\mu_1} \gamma_{\mu_2} \gamma_{\mu_3} P_L b)
                  \sum_q (\bar{q} \gamma^{\mu_1} \gamma^{\mu_2} \gamma^{\mu_3} q)\;, \nonumber\\
\mathcal{O}_6& = (\bar{s} \gamma_{\mu_1} \gamma_{\mu_2} \gamma_{\mu_3} T^a P_L b)
                  \sum_q (\bar{q} \gamma^{\mu_1} \gamma^{\mu_2} \gamma^{\mu_3} T^a q)\;,\nonumber\\
\mathcal{O}_7& = \frac{e}{(4\pi)^2} m_b (\overline{s} \sigma^{\mu\nu} P_R b) F_{\mu\nu} \;, \nonumber\\
\mathcal{O}_8& = \frac{g}{(4\pi)^2} m_b (\bar{s} \sigma^{\mu \nu} T^a P_R b) G_{\mu \nu}^a \;,  \nonumber\\
\mathcal{O}_9& =  \frac{e^2}{(4\pi)^2} (\overline{s} \gamma^\mu P_L b) (\bar{\ell} \gamma_\mu \ell) \;,   \nonumber\\
\mathcal{O}_{10}& =  \frac{e^2}{(4\pi)^2} (\overline{s} \gamma^\mu P_L b) (\bar{\ell} \gamma_\mu \gamma_5 \ell) \;.
\end{align}
The same set of operators applies for $b \to d$ processes, with the valence strange quark substituted by a valence down quark. We have denoted the $b$-quark mass by $m_b$, the strong coupling by $g$, the SU(3)$_c$ generators by $T^a$, and the photon and gluon field-strength tensors by $F_{\mu\nu}$ and $G_{\mu \nu}^a$. The sums run over the relevant quark flavours $q=u, d, s, c, b$.

Global statistical fits of the \WCs with flavour physics data are a standard way to uncover evidence for possible beyond-\sm (BSM) physics contributions, in a way that remains agnostic to the precise high-scale theory that supersedes the \sm.

In our analysis we allow for modification of three Wilson coefficients:
\begin{align*}
C_7&=C_7^{\rm SM} + \Delta C_7\;,   \nonumber\\
C_9&=C_9^{\rm SM} + \Delta C_9\;,   \nonumber\\
C_{10}&=C_{10}^{\rm SM} + \Delta C_{10}\;,
\end{align*}
where $C_i^{\rm SM}$ is the SM value of the $i$th \WC ($\Re(C_{7,9,10}^{\rm SM}) = -0.29,4.20,-4.06$), whereas $\Delta C_i$ is its modification by some high-energy new physics. In the fit that we perform in this paper, we vary the real parts of $\Delta C_7$, $\Delta C_{9}$ and $\Delta C_{10}$ to best match the experimental results.

\section{Observables included in the fit}
\label{sec:exp}

In this section we will discuss the theoretical calculations of the observables that are included in the fit. We perform the calculations with the latest version of \flavbit \cite{flavbit}, which uses \superisofour \cite{Mahmoudi:2007vz,Mahmoudi:2008tp,Mahmoudi:2009zz}.  Below, we provide a brief description for completeness.

\subsection{Angular distribution and branching fraction of \BdToKstarmumu decays}

The decay $\HepParticle{B}{d}{} \to \PKstarzero \APmuon \Pmuon$ is of particular interest as it presents a wide variety of experimentally-accessible observables. On the other hand, in general the hadronic uncertainties in the theoretical predictions are large. The decay with $K^*$ on the mass shell has a 4-fold differential distribution
\begin{equation}
  \frac{d^4\Gamma[\HepParticle{B}{d}{} \to \PKstarzero(\to \PK \Ppi)\APmuon\Pmuon]}
       {d q^2\, d{\ensuremath{\cos{\theta_l}}\xspace}, d{\ensuremath{\cos{\theta_K}}\xspace}\, d\phi} =
  \frac{9}{32\pi} \sum_i J_i(q^2)\, g_i(\theta_l, \theta_K, \phi)\,,
\end{equation}
with respect to the three angles $\theta_l$, $\theta_K$, and $\phi$ (as defined in~\cite{Egede:2008uy}) and the dilepton invariant mass $q^2$.
In the low-$q^2$ region (where $q^2$ is below the $J/\psi$ resonance), the description of this decay is provided by the method of QCD-improved factorisation (QCDf) and the Soft-Collinear Effective Theory (SCET).

The functions $J_{1-9}$ can be written in terms of the transversity amplitudes, $A_0$, $A_\parallel$, $A_\perp$, $A_t$ (and $A_S$ if scalar operators are also considered and lepton mass is not neglected) \cite{Egede:2010zc}:
\begin{subequations}
\begin{align}
\label{eq:J-TA}
  J_1^s & = \frac{(2+\beta_\ell^2)}{4} \left[|{A_\perp^L}|^2 + |{A_\parallel^L}|^2 + (L\to R) \right]
            + \frac{4 m_\ell^2}{q^2} \text{Re}\left({A_\perp^L}^{}{A_\perp^R}^* + {A_\parallel^L}^{}{A_\parallel^R}^*\right)\,,
\\
  J_1^c & =  |{A_0^L}|^2 +|A_0^R|^2  + \frac{4m_\ell^2}{q^2}
               \left[|A_t|^2 + 2\text{Re}({A_0^L}^{}{A_0^R}^*) \right] + \beta_\ell^2 |A_S|^2 \,,
\\
  J_2^s & = \frac{ \beta_\ell^2}{4}\left[ |{A_\perp^L}|^2+ |{A_\parallel^L}|^2 + (L\to R)\right]\,,
\\
  J_2^c & = - \beta_\ell^2\left[|{A_0^L}|^2 + (L\to R)\right]\,,
\\
  J_3 & = \frac{1}{2}\beta_\ell^2\left[ |{A_\perp^L}|^2 - |{A_\parallel^L}|^2  + (L\to R)\right]\,,
\\
  J_4 & = \frac{1}{\sqrt{2}}\beta_\ell^2\left[\text{Re} ({A_0^L}^{}{A_\parallel^L}^*) + (L\to R)\right]\,,
\\
  J_5 & = \sqrt{2}\beta_\ell\left[\text{Re}({A_0^L}^{}{A_\perp^L}^*) - (L\to R)
- \frac{m_\ell}{\sqrt{q^2}}\, \text{Re}({A_\parallel^L} {A_S^*}+{A_\parallel^R} {A_S^*})
\right]\,,
\\
  J_6^s  & = 2\beta_\ell\left[\text{Re} ({A_\parallel^L}^{}{A_\perp^L}^*) - (L\to R) \right]\,,
\\
   J_6^c  &  =
 4 \beta_\ell  \frac{m_\ell}{\sqrt{q^2}}\, \text{Re} \left[ {A_0^L} {A_S^*} + (L\to R) \right]\,,
\\
  J_7 & = \sqrt{2} \beta_\ell \left[\text{Im} ({A_0^L}^{}{A_\parallel^L}^*) - (L\to R)
+ \frac{m_\ell}{\sqrt{q^2}}\, {\text{Im}}({A_\perp^L} {A_S^*}+{A_\perp^R} {A_S^*})
\right]\,,
\\
  J_8 & = \frac{1}{\sqrt{2}}\beta_\ell^2\left[\text{Im}({A_0^L}^{}{A_\perp^L}^*) + (L\to R)\right]\,,
\\
  J_9 & = \beta_\ell^2\left[\text{Im} ({A_\parallel^L}^{*}{A_\perp^L}) + (L\to R)\right]\,,
\end{align}
\end{subequations}
where $\beta_\ell \equiv \sqrt{1-4m_\ell^2/q^2}$ and $(L\to R)$ indicates the same terms as immediately preceding, but with $L$ and $R$ superscripts exchanged.

The transversity amplitudes are related to the \WCs and form factors as
 \begin{align}
A_{\perp}^{L,R}  &=  N \sqrt{2\lambda}  \bigg[
\left[ (C_9 + Y(q^2) + C_9^{\prime}) \mp (C_{10} + C_{10}^{\prime}) \right] \frac{ V(q^2) }{ M_B + M_V}
 + \frac{2m_b}{q^2} (C_7^{\rm eff} + C_7^{\prime}) T_1(q^2)
\bigg]\,, \\
A_{\parallel}^{L,R}  & = - N \sqrt{2}(M_B^2 - M_V^2) \bigg[ \left[ (C_9 + Y(q^2) - C_9^{\prime}) \mp (C_{10} - C_{10}^{\prime}) \right]
\frac{A_1(q^2)}{M_B-M_V}
\nonumber\\
& \qquad +\frac{2 m_b}{q^2} (C_7^{\rm eff} - C_7^{\prime}) T_2(q^2)
\bigg]\,,\\
A_{0}^{L,R}  &=  - \frac{N}{2 M_V \sqrt{q^2}}  \bigg\{
 \left[ (C_9 + Y(q^2) - C_9^{\prime}) \mp (C_{10} - C_{10}^{\prime}) \right]
\nonumber\\
 & \qquad \times
\bigg[ (M_B^2 - M_V^2 - q^2) ( M_B + M_V) A_1(q^2)
 -\lambda \frac{A_2(q^2)}{M_B + M_V}
\bigg]
\nonumber\\
& \qquad + {2 m_b}(C_7^{\rm eff} - C_7^{\prime}) \bigg[
 (M_B^2 + 3 M_V^2 - q^2) T_2(q^2)
-\frac{\lambda}{M_B^2 - M_V^2} T_3(q^2) \bigg]
\bigg\}\,, \\
 A_t  &= \frac{N}{\sqrt{q^2}}\sqrt{\lambda} \left[ 2 (C_{10} - C_{10}^{\prime}) + \frac{q^2}{m_\ell (m_b +m_q)} (C_{Q_2} - C_{Q_2}^\prime)  \right] A_0(q^2) \,,\\
\label{3.50}
 A_S  &= - \frac{2N}{m_b+m_q} \sqrt{\lambda} (C_{Q_1} - C_{Q_1}^\prime)  A_0(q^2) \,,
\end{align}
where $M_V$ is the $K^*$ (vector) meson mass, $m_q$ the spectator quark mass and
\begin{equation}
N =  V_{tb}^{\vphantom{*}}V_{ts}^* \left[\frac{G_F^2 \alpha^2}{3\cdot 2^{10}\pi^5 M_B^3}
 q^2  \beta_\ell \sqrt{\lambda(M_B^2,M_V^2,q^2)} \right]^{1/2}\,,
\end{equation}
with the K\"all\'en function
\begin{equation}
 \lambda(x,y,z) \equiv x^2  + y^2 + z^2 - 2 (xy+ yz  + xz)\,.\label{eq:lambda}
\end{equation}
$C_7^{\rm eff}$ is defined as
\begin{equation}
C_7^{\rm eff} = C_7(\mu)
-\frac{1}{3} C_3(\mu) -\frac{4}{9} C_4(\mu)
-\frac{20}{3} C_5(\mu) -\frac{80}{9} C_6(\mu)\,,\label{eq:C7eff}
\end{equation}
and the function $Y(q^2)$ is given by
\begin{align}
Y(q^2) &= h(q^2,m_c) \left( \frac{4}{3} C_1 + C_2 + 6 C_3 + 60 C_5\right)
-\frac{1}{2}h(q^2,m_b^{\rm pole}) \left( 7 C_3 + \frac{4}{3}C_4 + 76 C_5
  + \frac{64}{3} C_6\right)
\nonumber\\
& \quad-\frac{1}{2}h(q^2,0) \left( C_3 + \frac{4}{3}C_4 + 16 C_5
  + \frac{64}{3} C_6\right)
  + \frac{4}{3} C_3 + \frac{64}{9} C_5 + \frac{64}{27} C_6\,,
\end{align}
with
\begin{equation}
h(q^2,m_q) = -\frac{4}{9}\, \left( \ln\,\frac{m_q^2}{\mu^2} - \frac{2}{3}
- z \right) - \frac{4}{9}\, (2+z) \sqrt{|z-1|} \times
\left\{
\begin{array}{l@{\quad}l}
\displaystyle\arctan\, \frac{1}{\sqrt{z-1}} & z>1\\[10pt]
\displaystyle\ln\,\frac{1+\sqrt{1-z}}{\sqrt{z}} - \frac{i\pi}{2} & z \leq 1
\end{array}
\right.
\end{equation}
where $z=4 m_q^2/q^2$.
For the form factors $A_{1,2,3}(q^2)$, $V(q^2)$, $T_{1,2,3}(q^2)$ we use the combined LCSR+lattice results from Ref.\ \cite{Straub:2015ica}. The precise values of these form factors are correlated, and together depend on 21 nuisance parameters. When obtaining the correlation matrix between the theoretical uncertainties on the different observables that enter our fit, we include these parameters in our marginalisation over theoretical uncertainties using Monte Carlo methods in \superisofour, for each combination of \WCs.

In addition, the transversity amplitudes receive corrections arising from the hadronic part of the Hamiltonian, through the emission of a photon which itself turns into a lepton pair.  The leading contributions at low $q^2$ can be calculated within the QCD factorisation approach where an expansion of $\Lambda/m_b$ is employed, but the subleading nonfactorisable power corrections are difficult to estimate. The corrections to the transversity amplitudes can be written as
 \begin{align}
\delta A_{\perp}^{L,R}  &=  \frac{32 \pi^2 N m_B^3}{\sqrt2\, q^2} \left( \mathcal{N}_+(q^2) -  \mathcal{N}_-(q^2) \right)\,,\\
\delta A_{\parallel}^{L,R}  &=  \frac{32 \pi^2 N m_B^3}{\sqrt2\, q^2} \left( \mathcal{N}_+(q^2) + \mathcal{N}_-(q^2) \right)\,,\\
\delta A_{0}^{L,R}  &=  \frac{32 \pi^2 N m_B^3}{q^2} \left( \mathcal{N}_0(q^2) \right)\,.
 \end{align}
The QCDf contributions to $\mathcal{N}_{\lambda}(q^2)$ are
\begin{align}
\mathcal{N}^{\rm QCDf}_\pm &= -\frac{1}{16\pi^2} \frac{m_b}{m_B}\left[ (m_B^2 - m_V^2)\frac{2 E_V}{m_B^3} \left(\mathcal{T}_\perp^{-(t),{\rm nf+WA}} + \hat{\lambda}_u \mathcal{T}_\perp^{-(u)} \right)\right. \\
&\quad\left.\mp \frac{\sqrt{\lambda}}{m_B^2}\left(\mathcal{T}_\perp^{+(t),{\rm nf+WA}} + \hat{\lambda}_u \mathcal{T}_\perp^{+(u)} \right)
\right]\,,\nonumber\\
\mathcal{N}^{\rm QCDf}_0 &= -\frac{1}{16\pi^2} \frac{m_b}{m_B} \frac{\sqrt{q^2}}{2m_V} \left\{ \left[ (m_B^2 + 3 m_V^2 - q^2)\frac{2 E_V}{m_B^3} - \frac{\lambda}{(m_B^2 - m_V^2)m_B^2}\right] \right. \\
&\quad\left. \times \left(\mathcal{T}_\perp^{-(t),{\rm nf+WA}} + \hat{\lambda}_u \mathcal{T}_\perp^{-(u)} \right) - \frac{\lambda}{(m_B^2 - m_V^2)m_B^2} \left(\mathcal{T}_\parallel^{-(t),{\rm nf+WA}} + \hat{\lambda}_u \mathcal{T}_\parallel^{-(u)} \right)\right\}\,\nonumber,
\end{align}
where $\hat{\lambda}_u=(V_{ub}^{\vphantom{*}}V_{us}^*)/(V_{tb}^{\vphantom{*}}V_{ts}^*)$ and the expressions for $\mathcal{T}^\pm$ can be found in Ref.~\cite{Beneke:2004dp}. The remaining hadronic corrections are unknown, and are assumed to be a fraction of the leading order non-factorisable contribution. They can be parameterised by multiplying $Y(q^2)$ and the $\delta A_{\lambda}^{L,R}$ by
\begin{equation}
 \left[1 + a_\lambda^{L,R} + b_\lambda^{L,R} \left(\frac{q^2}{6\;{\rm GeV}^2/c^4}\right)\right]\,,\label{eq:parameterisation}
\end{equation}
where $a_\lambda^{L,R}$ and $b_\lambda^{L,R}$ are taken as uncorrelated complex nuisance parameters.  We model the distributions of their amplitudes as Gaussians centered at 0, with variances of 10\% and 25\% respectively in the low-$q^2$ region. In the high $q^2$ region, we assign a variance of 10\% to the distribution of the amplitude of $a_\lambda^{L,R}$, and neglect the term proportional to $b_\lambda^{L,R}$, as in this regime the relative variation of $q^2$ is small compared to its variation at low $q^2$, so one can neglect higher orders in the $q^2$ expansion (and the sensitivity to New Physics at high $q^2$ is very small anyway). The phases are unknown constants (see Refs.~\cite{Hurth:2016fbr,Chobanova:2017ghn} for more details).

The traditional set of observables used to probe \BdToKstarmumu decays consists of the differential branching fraction
\begin{equation}
\frac{d\Gamma}{dq^2} = \frac{3}{4} \bigg( J_1 - \frac{J_2}{3} \bigg)\,,
\end{equation}
where $J_i \equiv 2 J^s_i + J^c_i$, and the angular observables
\begin{align}
F_L(q^2) &\equiv \frac{|{A}_0|^2}{|{ A}_0|^2 + |{A}_{\parallel}|^2+ |A_\perp|^2}\,,\\
A_{\rm FB}(q^2) & \equiv \int_{-1}^0 d\cos\theta_l\, \frac{d^2\Gamma}{dq^2 \, d\cos\theta_l} \Bigg/\frac{d\Gamma}{dq^2} - \int_{0}^1 d\cos\theta_l\, \frac{d^2\Gamma}{dq^2 \, d\cos\theta_l} \Bigg/\frac{d\Gamma}{dq^2}\nonumber\\
&= \frac{3}{8} J_6 \Bigg/ \frac{d\Gamma}{dq^2}\,.
\end{align}

In order to minimise the hadronic uncertainties emerging from form factor contributions to the $\HepParticle{\PB}{d}{} \to \PKstarzero \APmuon\Pmuon$ decay, angular observables have been constructed offering specific form-factor-independent observables (at leading order) \cite{Matias:2012xw,Egede:2008uy,Egede:2010zc,Descotes-Genon:2013vna}. One such set of observables is the so-called \emph{optimised observable set}, $P_i^\prime$, defined as

\begin{align}
\av{P_1}_{\rm bin}&= \frac12 \frac{\int_{{\rm bin}} dq^2 [J_3+\bar J_3]}{\int_{{\rm bin}} dq^2 [J_{2s}+\bar J_{2s}]}\ ,
&\av{P_2}_{\rm bin} &= \frac18 \frac{\int_{{\rm bin}} dq^2 [J_{6s}+\bar J_{6s}]}{\int_{{\rm bin}} dq^2 [J_{2s}+\bar J_{2s}]}\ ,\nonumber\\
\av{P'_4}_{\rm bin} &= \frac1{{\cal N}_\bin^\prime} \int_{{\rm bin}} dq^2 [J_4+\bar J_4]\ ,
&\av{P'_5}_{\rm bin} &= \frac1{2{\cal N}_\bin^\prime} \int_{{\rm bin}} dq^2 [J_5+\bar J_5]\ ,\nonumber\\
\av{P'_6}_{\rm bin} &= \frac{-1}{2{\cal N}_\bin^\prime} \int_{{\rm bin}} dq^2 [J_7+\bar J_7]\ ,
& \av{P'_8}_{\rm bin} &= \frac{-1}{{\cal N}_\bin^\prime} \int_{{\rm bin}} dq^2 [J_8+\bar J_8]\ ,
\label{Pi}
\end{align}
where the normalisation ${\cal N}_\bin^\prime$ is given by
\begin{equation}
{\cal N}_\bin^\prime = {\textstyle \sqrt{-\int_\bin dq^2 [J_{2s}+\bar J_{2s}] \int_{{\rm bin}} dq^2 [J_{2c}+\bar J_{2c}]}}\ .
\end{equation}

Alternatively, one can define the observables~\cite{Sinha:1996sv,Kruger:1999xa,Altmannshofer:2008dz}
\begin{equation}
S_i = \dfrac{J_{i(s,c)}+\bar{J}_{i(s,c)}}{\frac{d\Gamma}{dq^2}+\frac{d\bar{\Gamma}}{dq^2}}\,,
\end{equation}
which are related to the $P_i$ set as $S_{i} = P'_{i}\,\sqrt{F_L(1-F_L)}$.

The most important measurements for the interpretation of \bTosmumu transitions in terms of new physics are the angular observables in various \qsq bins of the \BdToKstarmumu decay. They are currently measured by four collaborations: LHCb, Belle, ATLAS and CMS, with the most recent measurement being an LHCb analysis of part of the LHC Run II dataset~\cite{Aaij:2020nrf}. In the case of this measurement, the whole set of angular observables is available with the full correlation matrix. In particular, LHCb provide{\color{blue}s} angular observables in the $S_i$ basis~\cite{Sinha:1996sv,Kruger:1999xa,Altmannshofer:2008dz} as well as the so-called optimised observables~\cite{Matias:2012xw}. The optimised observables are ``clean'' from hadronic uncertainties only at leading order, and with the current precision this is not enough for phenomenological applications. Furthermore, the optimised observables are non-linearly correlated with each other. In the following, we therefore use the measurements in the $S_i$ basis. As was pointed out in~\cite{Gratrex:2015hna}, the conventional theoretical and experimental angular observables differ by a minus sign in the case of $S_4$, $S_7$ and $S_9$, which we make sure to take into account.

The analyses of Belle \cite{Wehle:2016yoi}, CMS \cite{Sirunyan:2017dhj} and ATLAS \cite{Aaboud:2018krd} include measurements of only a subset of the angular observables. This is due to the fact that their datasets contain a smaller number of \BdToKstarmumu decays than that of LHCb, so the full angular distributions cannot be determined without folding some of the angles~\cite{Aaij:2013iag}. In our fit, we use all observables for which measurements by Belle, CMS or ATLAS are currently available.

In addition to the angular observables, we use the measured branching fraction of the \BdToKstarmumu decay in various \qsq bins. Currently the only measurement that distinguishes the $s$-wave and $p$-wave contributions is the LHCb one~\cite{Aaij:2016flj}. The theoretical framework discussed in \secref{sec:theory} can only describe the $p$-wave contribution. It is therefore of crucial importance to take into account the branching fraction measurement that subtracts the $s$-wave contribution.

In addition to the \BdToKstarmumu observables, we also include the corresponding observables from decays with electrons in our fit, i.e. \BdToKstaree.

\subsection{Branching fraction of \BsTophimumu decays}
\label{sec:phimumuBR}
The decay \BsTophimumu is also a \bTosmumu transition, but with a valence strange quark. The decay has only currently been measured by the LHCb collaboration~\cite{Aaij:2015esa}. In contrast to the \BdToKstarmumu decay, the \BsTophimumu decay is not self-tagging, and thus there is no experimental access to the most relevant $CP$-averaged observables $S_i$. Therefore, we use only the branching fraction information in our analysis. Because the $\Pphi$ meson has a much narrower width than the \PKstar meson, the $s$-wave pollution is negligible in this case.

The calculations for this decay are very similar to the ones for \BdToKstarmumu, with the main difference being that the spectator quark is a strange quark, and the meson masses and form factors are different. Here we use the form factors from the LCSR+lattice results \cite{Straub:2015ica}.

Because the $\HepParticle{B}{s}{} \to \Pphi \APmuon\Pmuon$ decay is not self-tagging, the untagged average over the $\HepParticle{\bar{B}}{s}{}$ and $\HepParticle{B}{s}{}$ decay distributions is required.  Defining~\cite{Descotes-Genon:2015hea}
\begin{equation}
\widetilde{J}_i=\zeta_i \bar{J}_i\ ,
\end{equation}
with
\begin{equation}\label{eq:zetadef}
\zeta_i=1\quad{\rm for}\quad i=1s,1c,2s,2c,3,4,7\,; \qquad
\zeta_i=-1\quad{\rm for}\quad i=5,6s,6c,8,9\,,
\end{equation}
and
\begin{equation}
 x=\frac{\Delta M}{\Gamma}\,,\qquad y=\frac{\Delta \Gamma}{2\Gamma}\,,
\end{equation}
the averaged $J_i$ functions are computed for LHCb with
\begin{eqnarray}
\label{eq:<J+Jt>Had}
\langle J_i + \bar J_i\rangle_{\rm Hadronic}
 &=& \frac{1}{\Gamma} \left[\frac{J_i+\widetilde J_i}{1-y^2}-\frac{yh_j}{1-y^2}\right]\,,
\end{eqnarray}
and the time-dependent decay rate is given by
\begin{eqnarray}
\left\langle\frac{d\Gamma}{dq^2}\right\rangle &=&  \frac{1}{\Gamma(1-y^2)} \langle{\cal I}\rangle\ ,\\[2mm]
\langle{\cal I}\rangle _{\rm Hadronic}
&=&\frac{3}{4}\bigg[2(J_{1s}+\bar{J}_{1s}-y\,h_{1s})+(J_{1c}+\bar{J}_{1c}-y\,h_{1c}) \bigg] \nonumber\\
&&-\frac{1}{4}\bigg[2(J_{2s}+\bar{J}_{2s}-y\,h_{2s})+(J_{2c}+\bar{J}_{2c}-y\,h_{2c}) \bigg]\,,
\end{eqnarray}
where ${\cal I}$ is the usual normalisation considered in analyses of the angular coefficients.

The coefficients $h_i$ relevant for the decay rate are:
\begin{eqnarray}
h_{1s}&=&\frac{2+\beta_\ell^2}{2}{\rm Re}\left[e^{i\phi}\left(\widetilde{A}^{L}_{\perp}A^{L*}_{\perp}+\widetilde{A}^{L}_{||}A^{L*}_{||}+\widetilde{A}^{R}_{\perp}A^{R*}_{\perp}+\widetilde{A}^{R}_{||}A^{R*}_{||}\right)\right]\\
\nonumber &&\qquad  +\frac{4m_\ell^2}{q^2}{\rm Re}\left[e^{i\phi}\left(\widetilde{A}^{L}_{\perp}A^{R*}_{\perp}+\widetilde{A}^{L}_{||}A^{R*}_{||}\right)+e^{-i\phi}\left(A^{L}_{\perp}\widetilde{A}^{R*}_{\perp}+A^{L}_{||}\widetilde{A}^{R*}_{||}\right)\right]\,,\\
h_{1c}&=&2{\rm Re}\left[e^{i\phi}\left(\widetilde{A}^{L}_{0}A^{L*}_{0}+\widetilde{A}^{R}_{0}A^{R*}_{0}\right)\right]\\\nonumber
&&\qquad +\frac{8m_\ell^2}{q^2}\left\{{\rm Re}\left[e^{i\phi}\widetilde{A}_tA_t^*\right]+{\rm Re}\left[e^{i\phi}\widetilde{A}^{L}_0A^{R*}_0+e^{-i\phi}A^{L}_0\widetilde{A}^{R*}_0\right]\right\}+2\beta_\ell^2{\rm Re}\left[e^{i\phi}\widetilde{A}_SA_S^*\right]\,,\\
h_{2s}&=&\frac{\beta_\ell^2}{2}{\rm Re}\left[e^{i\phi}\left(\widetilde{A}^{L}_{\perp}A^{L*}_{\perp}+\widetilde{A}^{L}_{||}A^{L*}_{||}+\widetilde{A}^{R}_{\perp}A^{R*}_{\perp}+\widetilde{A}^{R}_{||}A^{R*}_{||}\right)\right]\,,\\
h_{2c}&=&-2\beta_\ell^2{\rm Re}\left[e^{i\phi}\left(\widetilde{A}^{L}_{0}A^{L*}_{0}+\widetilde{A}^{R}_{0}A^{R*}_{0}\right)\right]\,,
\end{eqnarray}
where $\phi = 2 \beta_s$, $\sin\phi = 0.04$, $x=27$ and $y=0.06$ \cite{PhysRevD.98.030001}. The amplitudes $\widetilde{A}_X$ denote the amplitudes $A_X(\bar{B}\to f)$ in which CP-conjugation is not applied to the final state.
The hadronic uncertainties due to power corrections are accounted for using Eq.~(\ref{eq:parameterisation}).

\subsection{Branching fraction of \BToKmumu decays}
\label{sec::kmumuBR}

Another member of the \bTosmumu transition family is the decay \BToKmumu. Because the $\PK$ is a scalar, the decay kinematics can be described with only one helicity angle, and the angular distribution has only two observables, which are in fact not sensitive to the \WCs that we consider here. We therefore consider only the branching fraction of the \BToKmumu decay. The decay was measured by LHCb~\cite{Aaij:2012vr} and the $B$-factories Babar~\cite{Aubert:2008ps} and Belle~\cite{Wei:2009zv} . In our fits, we only include data from LHCb, as the uncertainties of the $B$-factory measurements are more than a factor of 4 larger, and therefore do not contribute much to the global picture.

The $B\to K \ell \ell$ matrix element can be written as \cite{Bobeth:2007dw}
\begin{align}
  \label{eq:matrix:el}
  {\cal M}(B\to K \ell \ell) & = i \frac{G_F \alpha_e}{\sqrt{2} \pi} V_{tb}^{} V_{ts}^{\ast}\,\, 
     \Bigg( F_V\, p_B^{\mu}\, [\bar{\ell}\gamma_{\mu} \ell] + F_A\, p_B^{\mu}\,
     [\bar{\ell} \gamma_{\mu}\gamma_5 \ell] \\
  & \hspace{4.3cm} + (F_S + \cos \theta F_T) \, [\bar{\ell}\ell]
      + (F_P + \cos \theta F_{T5}) \, [\bar{\ell}\gamma_5 \ell] \Bigg)\,,
   \nonumber
\end{align}
where $\theta$ is the angle between $\ell^-$ and the flight direction of $\bar{B}$ in the dilepton rest frame.

The $F_i$ functions are defined as \cite{Becirevic:2012fy}
\begin{align}
  \label{eq:FV}
  F_V(q^2) & = (C_9 + Y(q^2) +C_9^\prime)f_+(q^2) + \frac{2 m_b}{M_B+M_K} \left( C_7^{\rm eff}+C_7^\prime +\frac{4 m_\ell}{m_b} C_T \right)f_T(q^2)\,,
  \\[4mm]
  F_A(q^2) & = (C_{10}+C_{10}^{\prime})f_+(q^2)\,,
  \\[4mm]
  F_S(q^2) & = \frac{M_B^2 - M_K^2}{2 (m_b -m_s)} (C_S+C_S^\prime) f_0(q^2)\,,
  \\[4mm]
  F_P(q^2) & = \frac{M_B^2 - M_K^2}{2 (m_b -m_s)} (C_P+C_P^\prime) f_0(q^2)  \\ \nonumber
               &- m_\ell (C_{10}+C_{10}^{\prime}) \left[ f_+(q^2) -\frac{M_B^2 - M_K^2}{q^2} \left( f_0(q^2)-f_+(q^2) \right) \right]\,,
  \\[4mm]
  F_T(q^2) & = \frac{2 \sqrt{\lambda}\, \beta_\ell}{M_B + M_K} C_T f_T(q^2)\,,
  \\[4mm]
  F_{T5}(q^2) & = \frac{2 \sqrt{\lambda}\, \beta_\ell}{M_B + M_K} C_{T5} f_T(q^2)\,.
\end{align}
where $C_T$ and $C_{T5}$ are tensor \WCs, which we take to be equal to zero in our analysis, and $f_0,f_+,f_-,f_T$ are form factors. We consider the LCSR+lattice results from \cite{Altmannshofer:2014rta} together with their uncertainties and correlations.

$F_V$ receives corrections from hadronic terms:
\begin{equation}
 \delta F_V = \frac{2 m_b}{M_B+M_K} \mathcal{T}_P\,,
\end{equation}
where $\mathcal{T}_P$ is given in \cite{Beneke:2001at,Bobeth:2007dw}.

To evaluate the uncertainties due to higher-order corrections, we again use the parameterisation
\begin{equation}
F_i \rightarrow F_i \left[1 + a_i + b_i \left(\frac{q^2}{6\;{\rm GeV}^2/c^4}\right)\right]\,,
\end{equation}
with $a_i$ and $b_i$ uncorrelated complex nuisance parameters.  As for the \BdToKstarmumu angular observables, following the prescription of~\cite{Hurth:2016fbr}, we model the distributions of their amplitudes as Gaussians centered at 0, with respective variances of 10\% and 25\% in the low-$q^2$ region, and 10\% and 0\% in the high-$q^2$ region, and take their phases to be uniformly random.

The decay rate is then given by
\begin{align}
 \Gamma (B \to K \ell^+ \ell^-) & = 2 \left(A_\ell + \frac{1}{3} C_\ell \right)\,,
\end{align}

where
\begin{align}
  A_\ell & = \int_{q^2_{\rm min}}^{q^2_{\rm max}} dq^2\, a_\ell(q^2)\,, &
  C_\ell & = \int_{q^2_{\rm min}}^{q^2_{\rm max}} dq^2\, c_\ell(q^2)\,.
\end{align}
with
\begin{align}
  \label{eq:al}
  a_\ell(q^2) & = {\cal{C}}(q^2) \Big[ q^2 \left( \beta^2_\ell |F_S|^2 + |F_P|^2 \right)
                   + \frac{\lambda}{4}  \left(|F_A|^2 + |F_V|^2\right) \\ \nonumber
              &\quad \quad + 2 m_\ell \left(M_B^2 - M_K^2 + q^2\right) {\rm Re}(F_P F_A^\ast) + 4 m_\ell^2 M_B^2 |F_A|^2 \Big] \,,\\[4mm]
  c_\ell(q^2) & ={\cal{C}}(q^2) \Big[ q^2 \left( \beta_\ell^2 |F_T|^2+ |F_{T5}|^2 \right) - \frac{\lambda}{4} \beta_\ell^2 \left(|F_A|^2 + |F_V|^2\right)
                  + 2 m_\ell \sqrt{\lambda} \beta_\ell {\rm Re}(F_{T} F_V^\ast)\Big]\,,
\end{align}
where $\lambda = \lambda(M_B^2, M_K^2, q^2)$ is as defined in Eq.~(\ref{eq:lambda}),
\begin{align}
 {\cal{C}}(q^2)= \Gamma_0\, \beta_\ell \, \sqrt{\lambda}\,,
\end{align}
with
\begin{align}
 \Gamma_0 &= \frac{G_F^2 \alpha_e^2 |V_{tb}^{} V_{ts}^{\ast}|^2}{512 \pi^5 M_B^3}\,.
\end{align}

\subsection{Branching fraction of \BTomumu decays}

The rare decay $\HepParticle{B}{s}{} \to \APmuon \Pmuon$ is strongly helicity-suppressed in the \sm and proceeds via $Z^0$ penguin and box diagrams, but can receive large contributions from BSM physics. The main contribution to this decay is from the effective operator $O_{10}$ in the SM and from the scalar and pseudoscalar operators $O_{S,P}$ in some BSM scenarios. As $O_{10}$ has no contamination from four-quark operators, the generalisation to $B_d$ decay is straightforward.

The branching fraction is given by
\begin{align}
\mathrm{BR}(\HepParticle{B}{s}{} \to \APmuon \Pmuon) &= \frac{G_F^2 \alpha^2}{64 \pi^3} f_{B_s}^2 \tau_{B_s} m_{B_s}^3 |V_{tb}V_{ts}^*|^2 \sqrt{1-\frac{4 m_\mu^2}{m_{B_s}^2}} \\
&\times \left[\left(1-\frac{4 m_\mu^2}{m_{B_s}^2}\right) \left| {\left(\frac{m_{B_s}}{m_b+m_s}\right)} C_S \right|^2 +  \left| {\left(\frac{m_{B_s}}{m_b+m_s}\right)} C_P + 2 \, C_{10} \frac{m_\mu}{m_{B_s}}  \right|^2\right] \;, \nonumber
\end{align}
where $f_{B_s}$ is the $B_s$ decay constant, $m_{B_s}$ is the $B_s$ meson mass and $\tau_{B_s}$ is the $B_s$ mean life.  As we consider only scenarios where new physics enters through modifications of $C_7$, $C_9$ and/or $C_{10}$, in our analysis we set $C_S = C_P = 0$.

The main theoretical uncertainty comes from $f_{B_s}$, which is determined with lattice QCD. We use  $f_{B_s} = 227.7$ MeV~\cite{Aoki:2019cca}. The main parametric uncertainty is from the CKM matrix element $V_{ts}$.

Within the minimal flavour violation approximation, the $B_{d} \to \ell^+ \ell^-$ rate can be obtained from the $B_{s} \to \ell^+ \ell^-$ rate simply by exchanging $s \to d$ in the above formula.

\subsection{Branching fraction of inclusive \bTosgamma decays}

Last but not least of the relevant processes in our study is the inclusive branching fraction of \bTosgamma. As an inclusive decay, it does not suffer from form factor uncertainties, and it therefore provides the strongest constraint on the $C_7$ \WC.

The branching fraction of $B\to X_s \gamma$ for a photon energy cut $E_\gamma > E_0$ is given by~\cite{Grigjanis:1988iq,Grinstein:1987vj,Misiak:2006zs,Misiak:2006ab,Czakon:2015exa,Misiak:2017woa,Misiak:2020vlo}
\begin{equation}
{\rm BR} (B \to X_s \gamma)_{E_\gamma > E_0}
=  {\rm BR}  (B \to X_c e \bar \nu)_{\rm exp} \, {6 \alpha_{\rm em} \over \pi C} \,
\left| V_{ts}^* V_{tb}^{}\over V_{cb}^{}\right|^2
\, \Big[ P(E_0) + N(E_0) \Big] \,,  \\
\end{equation}
where $\alpha_{\rm em} = \alpha_{\rm em}^{\rm on~shell}$~\cite{Czarnecki:1998tn}, $C = |V_{ub}|^2 / |V_{cb}|^2 \,\times\, \Gamma[B\to X_c e \bar \nu] / \Gamma [ B\to X_u e \bar \nu] $ and $P(E_0)$ and $N(E_0)$ denote the perturbative and nonperturbative contributions, respectively. We adopt the standard experimental cut $E_0=1.6$\,GeV.

The perturbative contributions are known at NNLO precision, while the nonperturbative corrections are estimated to be below $5\%$~\cite{Gunawardana:2019gep}.
The main sources of theoretical uncertainty are nonperturbative, parametric and perturbative (scale) uncertainties, and ambiguity arising from interpolation between results computed at different values of $m_c$.

The perturbative part of the Wilson coefficients is parameterised as
\begin{eqnarray}
P(E_0) = P^{(0)}(\mu_b) + \left(\frac{\alpha_s(\mu_b)}{4\pi}\right) \left[ P_1^{(1)}(\mu_b) + P_2^{(1)}(E_0,\mu_b) \right]  + {\cal O}\left(\alpha_s^2(\mu_b)\right)\,,
\end{eqnarray}
where bracketed superscripts indicate order in perturbation theory, and
\begin{eqnarray}
P^{(0)}(\mu_b) &=& \left[ C_7^{(0)\rm eff}(\mu_b)\right]^2\,, \nonumber\\
P_1^{(1)}(\mu_b) &=& 2\, C_7^{(0)\rm eff}(\mu_b) \,C_7^{(1)\rm eff}(\mu_b)\,,\nonumber\\
P_2^{(1)}(E_0,\mu_b) &=& \sum_{i,j=1}^{8} C_i^{(0)\rm eff}(\mu_b)\; C_j^{(0)\rm eff}(\mu_b) \; K_{ij}^{(1)}(E_0,\mu_b)\,.
\end{eqnarray}
The functions $K_{ij}^{(1)}$ can be found in Ref.~\cite{Misiak:2006ab}, $C_i^{\rm eff}(\mu) = C_i(\mu)$ for $i = 1, ..., 6$, $C_7^{\rm eff}(\mu)$ is given in Eq.~(\ref{eq:C7eff}), and
\begin{equation}
C_8^{\rm eff} = C_8(\mu) + C_3(\mu) -\frac{1}{6} C_4(\mu)
+ 20  C_5(\mu) -\frac{10}{3} C_6(\mu)\,.
\end{equation}

We also consider the branching fraction for \BtoKstarGamma\, in the fit, following the theory calculation in Ref.\ \cite{Beneke:2001at}.

\subsection{Other measurements}

Other potentially interesting experimental measurements of \bTosmumu decays are provided by observations of the decays of the $\PLambdab$ baryon, such as $\PLambdab \to \Lambda \mu \mu$. The LHCb collaboration has measured both the branching fraction~\cite{Aaij:2017ewm} and the angular distribution~\cite{Aaij:2018gwm} using the method of moments~\cite{Blake:2017une,Beaujean:2015xea}. We have not considered these measurements here as they have much larger uncertainties than those of the corresponding meson decays. It is worth pointing out, however, that the $\Lambda$ baryon is stable under strong interactions and therefore the computation of the form factors does not require a complicated treatment of multi-hadron states.  Once more experimental data  are available, recent~\cite{Detmold:2016pkz} and future developments of lattice calculations mean that this decay will eventually be placed on the same footing as other \bTosmumu transitions.

In the current analysis, we only consider the lepton-universal \WCs. Therefore, we do not include any of the observables explicitly designed to test violation of lepton flavour universality, such as $R_K$ or $R_{\PKstar}$. We defer the study of \WCs that violate lepton universality to future work.

\section{Statistical treatment}
\label{sec:stats}

We carry out global fits varying three \WCs: \recDSeven, \recDNine and \recDTen.  For each set of the three parameters, we compute the theoretical prediction for all considered observables and the covariance matrix corresponding to the uncertainties on the theoretical predictions arising from the variation of all theory nuisance parameters. We perform the fit using \gambit \cite{gambit,grev}, an open-source, modular package that combines theory calculations, experimental likelihoods, statistics and sampling routines. In particular, we use the \flavbit module \cite{flavbit} for computing all theoretical predictions and experimental likelihoods.  The latest version of \flavbit obtains observables via an interface to \superisofour~\cite{Mahmoudi:2007vz,Mahmoudi:2008tp,Mahmoudi:2009zz}, and likelihoods via an interface to the \heplike package~\cite{Bhom:2020bfe}, which retrieves experimental results and their correlated uncertainties from the \heplikedata repository \cite{HEPLikeData}.

Here we present only profile likelihood results, which we obtain via the interface in the \gambit \scannerbit module to the differential evolution sampler \diver \cite{ScannerBit}, run with a population of 20\,000 and a convergence threshold parameter of $10^{-5}$.  We also carried out an equivalent Bayesian analysis using the ensemble Markov Chain Monte Carlo sampler \twalk \cite{ScannerBit}; the results are practically identical to the profile likelihood ones, so we do not show them here.

To compute the theoretical covariance matrix, we follow a similar approach as the one described in \cite{Arbey:2016kqi}. Let us consider two observables $Q$ and $T$, which are subject to elementary sources of uncertainties, numbered $a=(1,\cdots,n)$. We denote the variations of the nuisance parameters as $\delta_a$, which have an impact on both observables. Assuming that the uncertainties are small enough to affect the observables linearly, the total variation of observable $Q$ is given at first order by:
\begin{equation}
 Q=Q_0\left(1+\sum_{a=1}^n\delta_a \Delta_Q^a\right)\,,
\end{equation}
where $\Delta_Q^a$ is the relative variance generated by the nuisance parameter $a$ and $Q_0$ is the central value.  We denote the covariance matrix between the nuisance parameters as
\begin{equation}
 \rho_{ab} = {\cov}[\delta_a , \delta_b]\,.
\end{equation}
such that the total relative variance of observable $Q$ is
\begin{equation}
 (\Delta_Q)^2 = \sum_{a,b} \rho_{ab} \Delta_Q^a \Delta_Q^b\,,
\end{equation}
the correlation coefficient between $Q$ and $T$ is
\begin{equation}
 (\Delta_{QT})^2 = \sum_{a,b} \rho_{ab}\, \Delta_{Q}^a \Delta_{T}^b\,,
\end{equation}
and the covariance matrix of observables $Q$ and $T$ is therefore
\begin{equation}
 \cov[Q,T] = \left(
    \begin{array}{cc}
        (\Delta_{Q})^2 (Q_0)^2 & (\Delta_{QT})^2 \, Q_0 T_0\\
        (\Delta_{QT})^2 \, Q_0 T_0 & (\Delta_T)^2 (T_0)^2
    \end{array}
\right)\,.
\label{eq::cov}
\end{equation}

In practice, most of the nuisance parameters are uncorrelated, so that $\rho_{ab} = \delta_{ab}$. The form factors on the other hand are strongly correlated, and we make sure to include their correlation matrices when computing Eq.\ \ref{eq::cov}.

In the following subsection we will discuss specifics of our treatment of different experimental likelihoods.

\subsection{Angular distribution of \BdToKstarmumu, \BdToKstaree and \BuToKstarmumu decays}
\label{sec:kstarmumu_ang_like}
The angular coefficients of \BdToKstarmumu, \BdToKstaree and  \BuToKstarmumu decays are measured by several experiments, using different methods and providing different information.

In the most recent LHCb publication~\cite{Aaij:2020nrf}, the angular observables are provided in bins of \qsq with the full experimental covariance matrix. In contrast to previous LHCb results~\cite{Aaij:2015oid}, the uncertainties provided are symmetric, which is a consequence of increased statistics. The constructed experimental likelihood has the form:
\begin{equation}
\ln\mathcal{L}(C_i)=-\frac{1}{2}V^{T}(C_i)\cov^{-1} (C_i) V(C_i)\,,
\label{eq:ndimgauss}
\end{equation}
where \l~denotes the likelihood, \cov~is the covariance matrix, and $V$ is the vector of differences between the measured values and the theory predictions for a given set of Wilson coefficients $(C_i)$.

In the case of other analyses~\cite{Abdesselam:2016llu,CMS:2017ivg,LHCb:2020gog}\footnote{In the case of the Belle experiment, we use the average between the muon and the electron mode}, the uncertainties are reported as asymmetric. In this case, we construct an experimental covariance matrix for each point depending on which of the asymmetric errors is relevant:
\begin{align}
\cov[Q,T]=
\begin{cases}
{\rm Corr}[Q,T]~\sigma^{Q}_+ \sigma^{T}_+, & \text{if } Q \geq Q_{obs} \text{ and }  T \geq T_{obs} \\
{\rm Corr}[Q,T]~\sigma^{Q}_+ \sigma^{T}_-, & \text{if } Q \geq Q_{obs} \text{ and }  T < T_{obs} \\
{\rm Corr}[Q,T]~\sigma^{Q}_- \sigma^{T}_+, & \text{if } Q < Q_{obs} \text{ and }  T \geq T_{obs} \\
{\rm Corr}[Q,T]~\sigma^{Q}_- \sigma^{T}_-, & \text{if } Q < Q_{obs} \text{ and }  T < T_{obs}\,, \\
\end{cases}
\label{eq:bifurgauss}
\end{align}
where $\sigma^{k}_{+}$, and $\sigma^{k}_{-}$ are the reported asymmetric uncertainties of the $k$th observable, which we take to be given by the sum in quadrature of the reported systematic and statistical uncertainties.  This is a refined treatment compared to some previous studies.

We then compute the total covariance matrix as the sum of the experimental and theoretical covariance matrices: $\cov=\cov_{\rm exp} + \cov_{\rm th}$.

\subsection{Branching fractions of the \BuToKstarmumu, \BdToKstarmumu, \BsTophimumu and \BToKmumu decays}

In addition to the angular observables, we also include likelihoods for the branching fractions of \BuToKstarmumu, \BdToKstarmumu, \BsTophimumu and \BToKmumu decays in our fit, in multiple \qsq bins. Currently only the LHCb collaboration has measured these observables~\cite{Aaij:2015esa,Aaij:2016flj,Aaij:2012vr}, with asymmetric uncertainties. We construct the likelihood in the same manner as in~\secref{sec:kstarmumu_ang_like}. The branching fractions of these decays are independent measurements and are statistically dominated. Therefore, no experimental correlation occurs between them. As for the \BdToKstarmumu angular observables, we take into account asymmetric uncertainties.

\subsection{Branching fractions of the \BsTomumu and \BdTomumu decays}

The branching fractions of \BsTomumu and \BdTomumu have both been measured by LHCb~\cite{Aaij:2017vad}, CMS~\cite{CMS:2019qnb} and ATLAS~\cite{Aaboud:2018mst}. All these measurements were simultaneous determinations of both the $B_d$ and $B_s$ modes. All three publications provide two-dimensional log-likelihood information, which we use to construct our likelihoods.

For each two-dimensional set of measurements, we profile over a two-dimensional Gaussian distribution $\mathcal{N}$ for the theoretical uncertainties on the branching ratios for \BsTomumu and \BdTomumu, giving a final likelihood
\begin{align}
\l(\mathrm{BR}_s, \mathrm{BR}_d) = \min_{\mathrm{BR}'_s, \mathrm{BR}'_d} \l_{\rm exp}(\mathrm{BR}'_s, \mathrm{BR}'_d)  \times \mathcal{N}( \mathrm{BR}'_s, \mathrm{BR}'_d | \mathrm{BR}_s, \mathrm{BR}_d, \cov)\,,
\end{align}
where $\l_{\rm exp}$ is the two-dimensional experimental likelihood, $\mathrm{BR}_s$ and $\mathrm{BR}_d$ are the theoretically-predicted branching fractions of  \BsTomumu and \BdTomumu decays respectively, while $\cov$ is the covariance matrix describing their correlated uncertainties.

As all three experimental results have similar sensitivities, we include all three in our total likelihood function.

\subsection{Inclusive branching fraction for \bTosgamma decays and exclusive branching fraction for \BtoKstarGamma~and \BtoKstaree}

We employ simple one-dimensional Gaussian likelihood based on the experimental measurement BR(\bTosgamma) = $\left(3.32\pm 0.15\right)\times 10^{-4}$, as recommended by the HFLAV collaboration~\cite{Amhis:2018udz}.  This value is based on a photon energy requirement of $E_{\gamma}>1.6$ GeV.

We do the same for \BtoKstarGamma, constructing a one-dimensional Gaussian likelihood based on the HFLAV recommendation: BR(\BtoKstarGamma) = $\left(4.1\pm 0.12\right)\times 10^{-5}$ \cite{Amhis:2018udz}.

The decay \BtoKstaree was measured by the LHCb Collaboration~\cite{LHCb:2020dof} in the low \qsq region of $\qsq \in \left[ 0.0008, 0.257 \right]$ GeV$^2$. The signal in this region is dominated by the contribution from the radiative electroweak penguin diagram, constraining $C_7$. The measurement consists of four amplitudes provided with correlations, from which we construct a four-dimensional likelihood function.

\begin{table}[t!]
\begin{center}
\begin{tabular}{|l|c|c|c|c|}
\hline
\WC & Best-Fit Point & 68.3\% interval & 95.4\% interval & 99.7\% interval\\ \hline \hline
\recDSeven &  $\phantom{-}0.011$ & $\left[-0.004, 0.025 \right]$ & $\left[-0.018, 0.038 \right]$ & $\left[-0.032, 0.050 \right]$ \\ \hline
\recDNine &  $-1.09\phantom{0}$ & $\left[-1.25, -0.93\right]$ & $\left[-1.39, -0.77\right]$ & $ \left[-1.52,-0.61\right]$  \\ \hline
\recDTen &  $\phantom{-}0.06\phantom{0}$ &  $\left[-0.09, 0.17 \right]$ & $\left[-0.22, 0.29\right]$ & $ \left[-0.34,0.41\right]$  \\ \hline
\end{tabular}
\end{center}
\caption{Results of the combined fit to the \recDSeven, \recDNine, \recDTen \WCs. For each \WC we give the best-fit value and the 1, 2 and 3$\sigma$ intervals.
\label{tab::WCresults}}
\end{table}

\begin{figure}
\hspace{-5mm}
\begin{tabular}{c@{}c@{}c}
\includegraphics[height=0.288\textwidth]{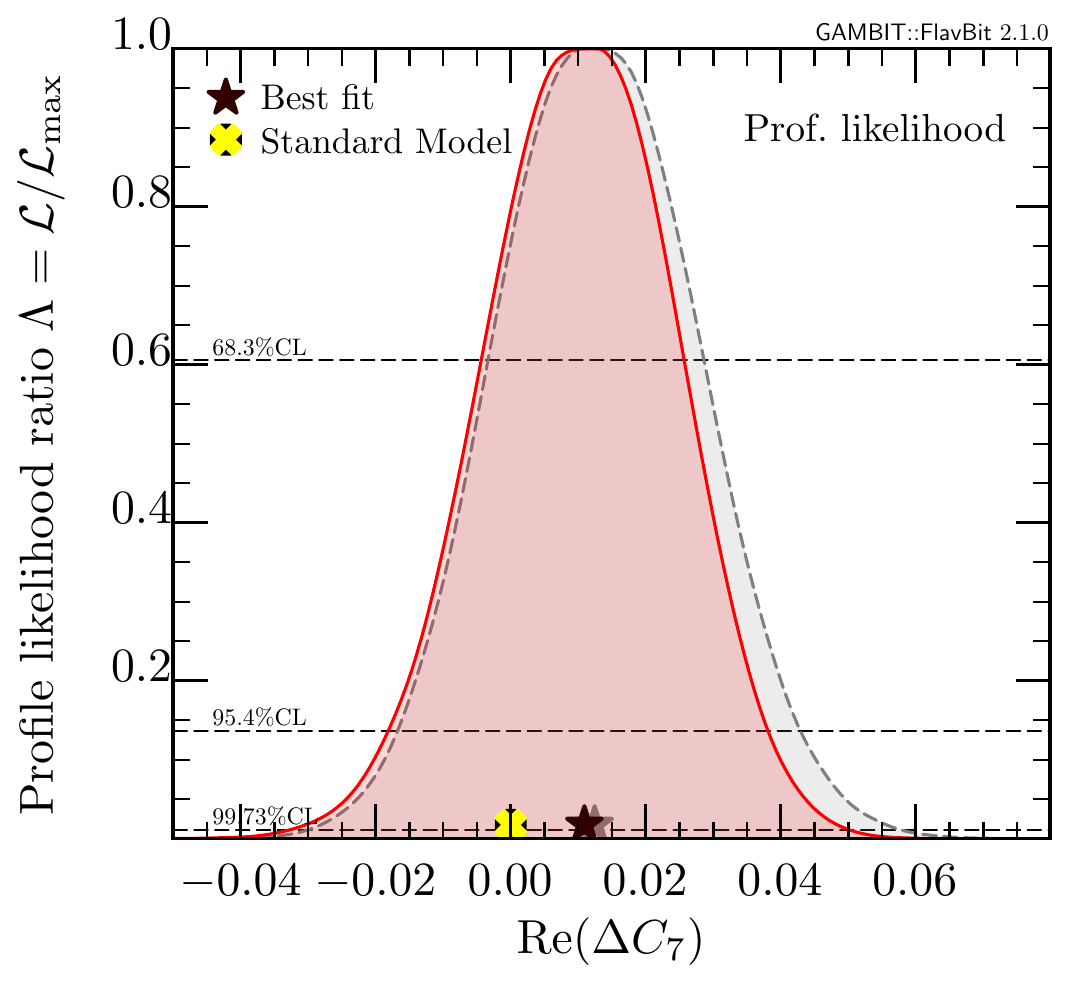}&~&~\\
\includegraphics[height=0.288\textwidth]{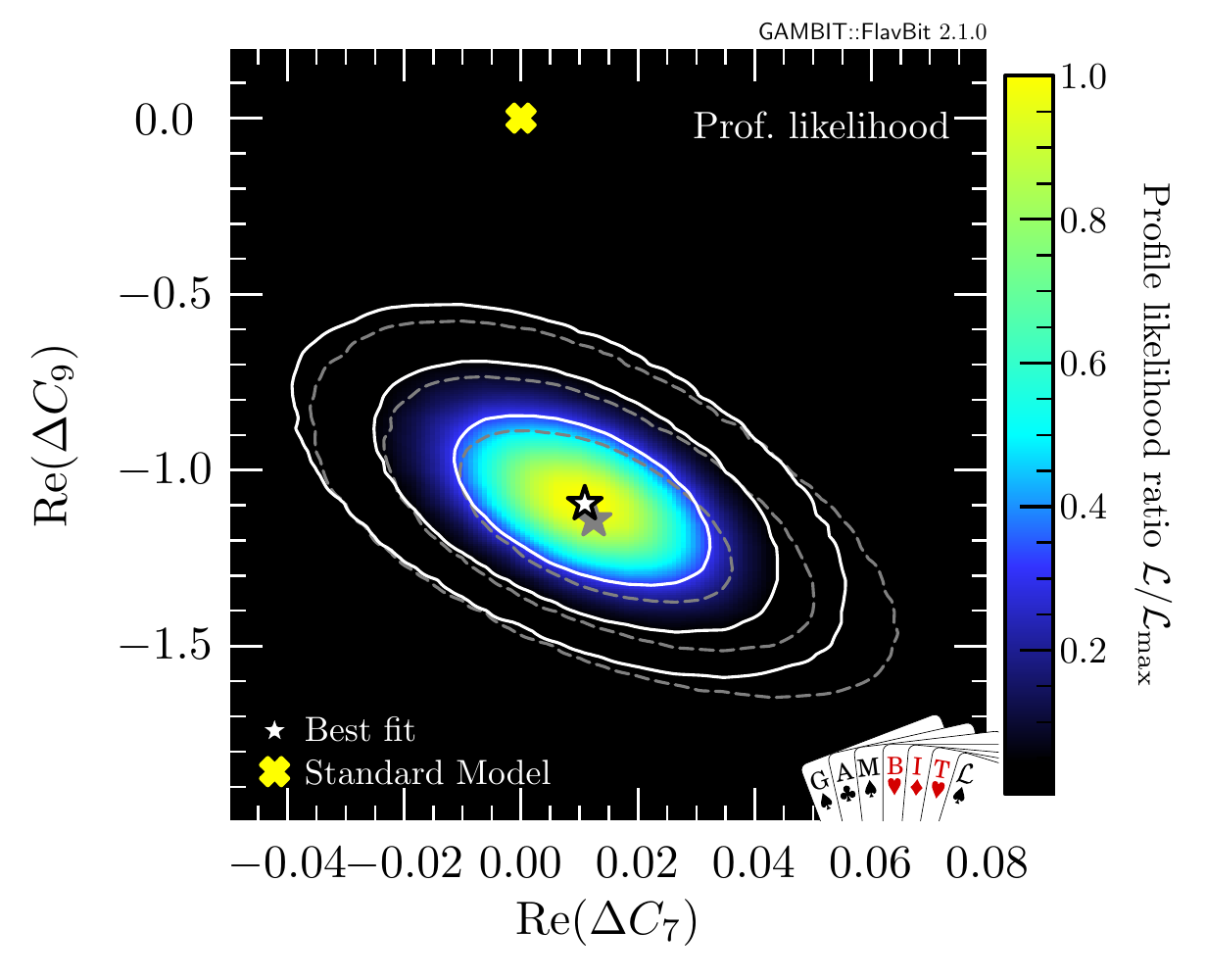}&
\includegraphics[height=0.288\textwidth]{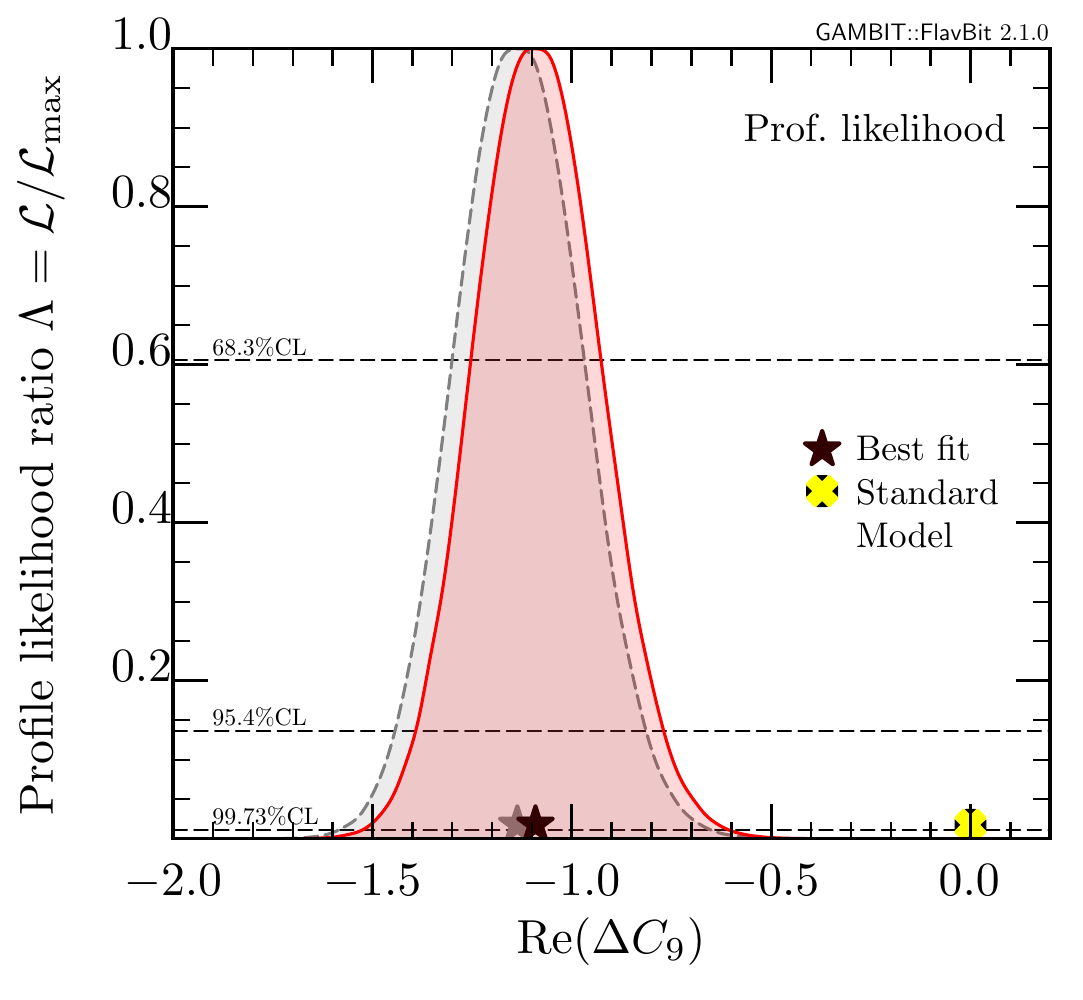}&~\\
\includegraphics[height=0.288\textwidth]{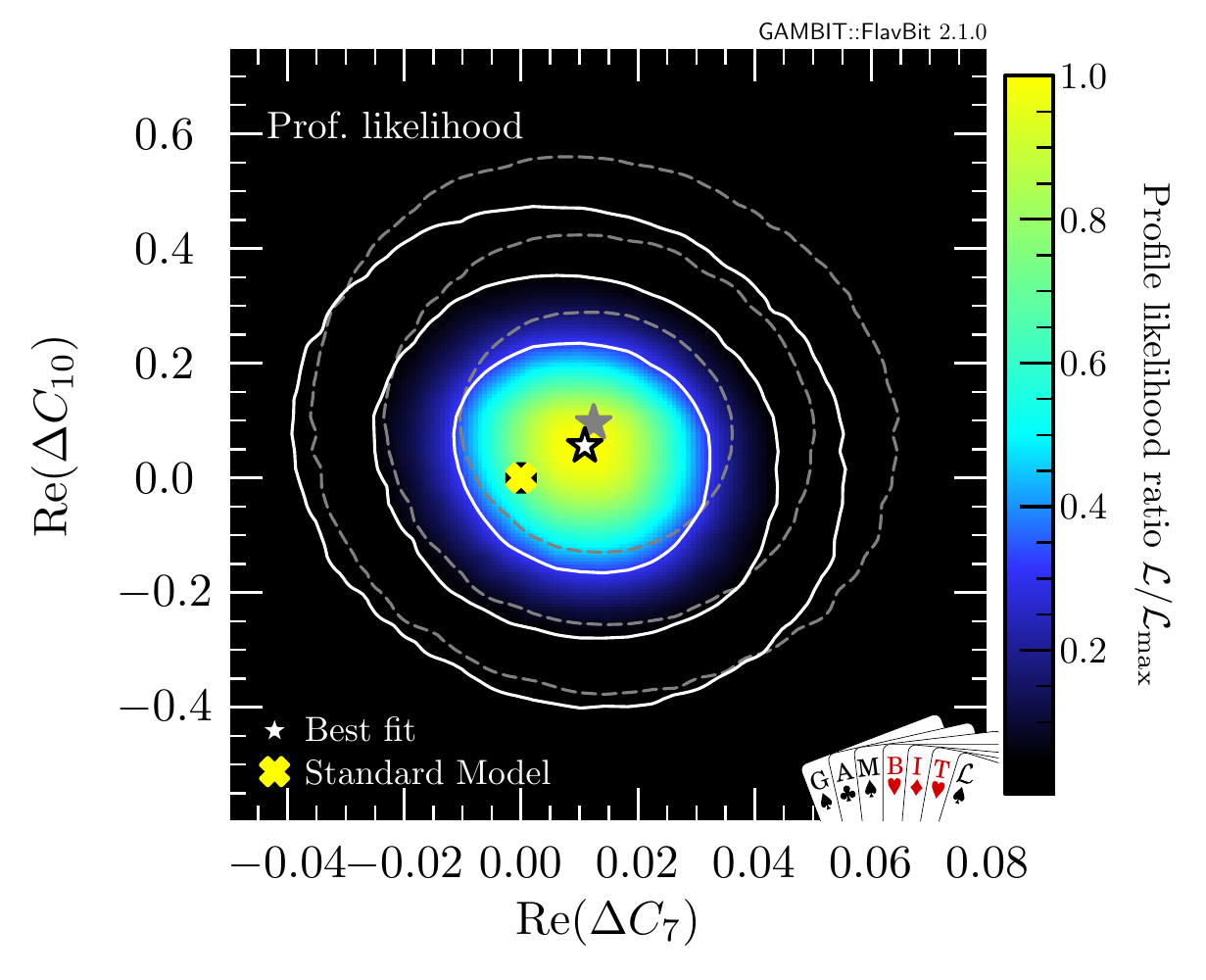}&
\includegraphics[height=0.288\textwidth]{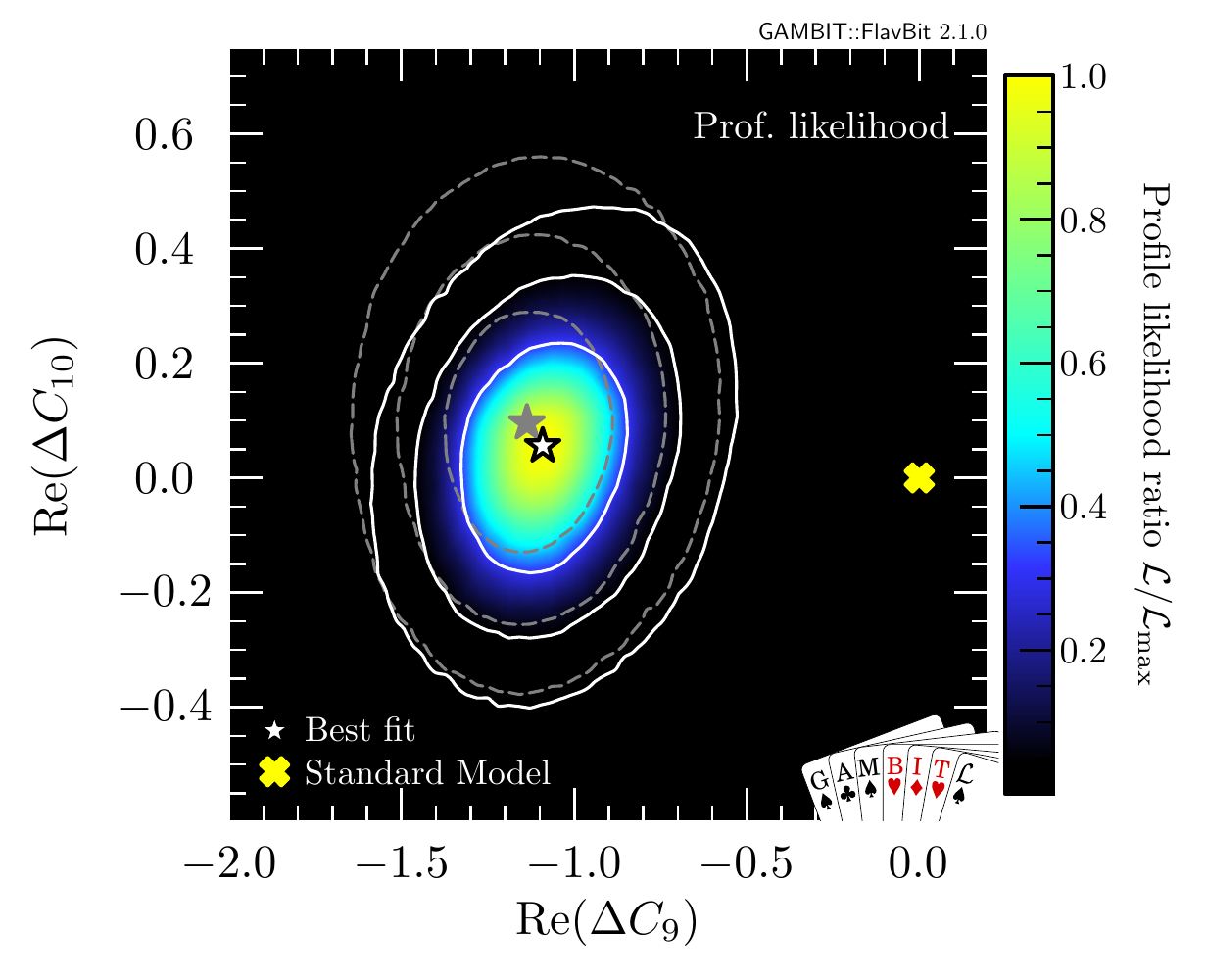}&
\includegraphics[height=0.288\textwidth]{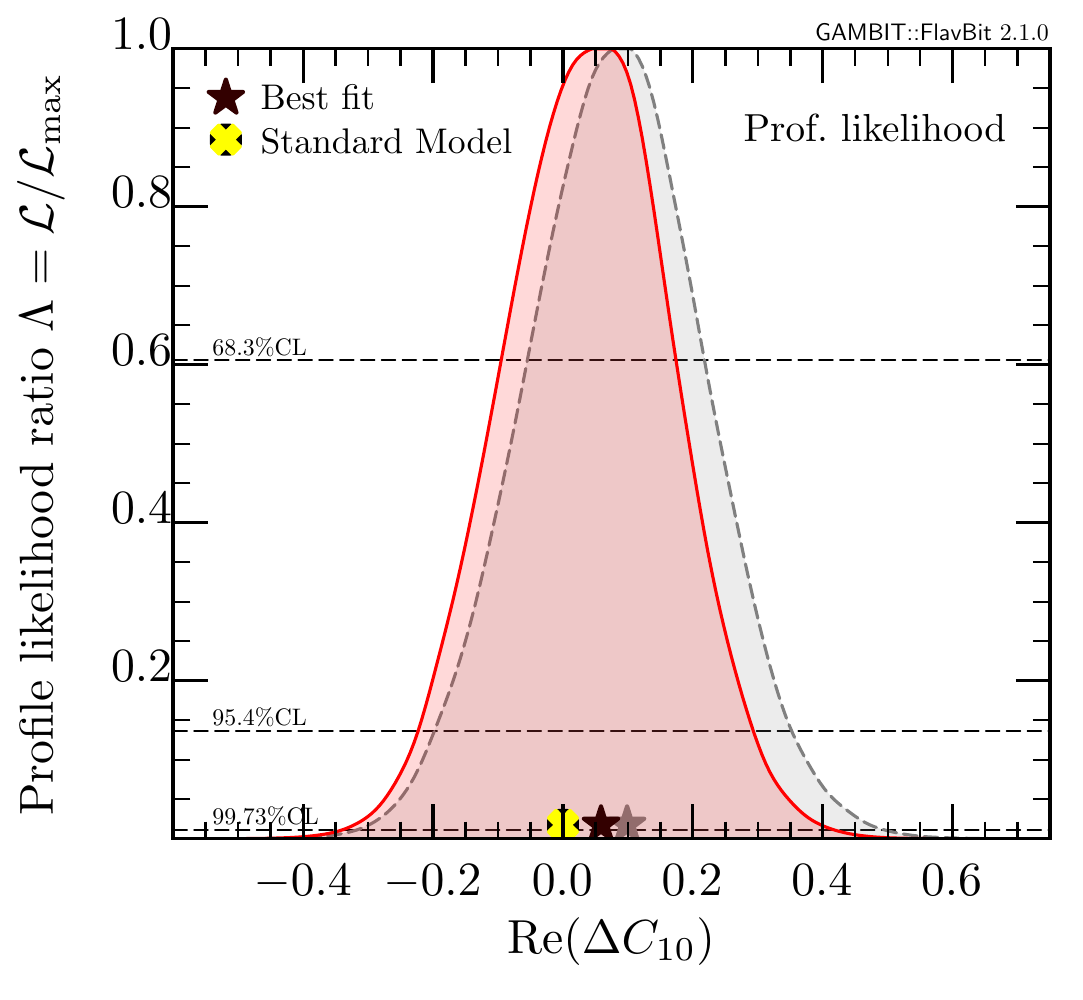}
\end{tabular}
\caption{Complete results of the combined fit to the \recDSeven, \recDNine and \recDTen \WCs, showing one- and two-dimensional profile likelihoods of each parameter.  Contour lines indicate 1, 2 and 3$\sigma$ confidence regions.  White contours and coloured shading in two-dimensional planes, and red one-dimensional curves, show our main results, where we compute theory covariances self-consistently for every combination of \WCs.  Grey contours and curves show the corresponding result when we approximate the theory covariance by its value in the Standard Model, across the entire parameter space.  The Standard Model prediction is indicated by a yellow cross.
\label{fig::WCresults}}
\end{figure}

\section{Results}
\label{sec:results}

\subsection{Current status}

In \tabref{tab::WCresults} and \figref{fig::WCresults} we present the main results of our global fit, providing one and two dimensional profile likelihoods for each of the \WCs. As can be seen, the strongest required modification to the \sm \WCs is in \recNine.  The best-fit points correspond to coupling strengths $\Re(C_7)/\Re(C_7^\text{SM})=0.96$, $\Re(C_9)/\Re(C_9^\text{SM})=0.74$, and $\Re(C_{10})/\Re(C_{10}^\text{SM})=0.99$ relative to the SM.  The agreement with the \sm can be quantified by comparing the log-likelihood of the best-fit point to that of the \sm.  This gives a total of $\Delta \ln\mathcal{L}=25.8$, which for the three degrees of freedom in our fit, corresponds to a $6.6\sigma$ exclusion of the \sm.
Considering just \recNine alone, i.e.\ also profiling out the impacts of allowing \recSeven and \recTen to vary, such that only a single degree of freedom remains, we find $\Delta \ln\mathcal{L}=19.6$.  This corresponds to a $6.3\sigma$ preference for a non-SM value of $C_{9}$.\footnote{These calculations assume that the asymptotic limit of Wilks' theorem holds, i.e.\ that in the asymptotic limit of a large data sample, twice the difference $\Delta \ln\mathcal{L}$ follows a $\chi^2$ distribution with $n$ degrees of freedom, where $n$ is the difference in dimensionality between the larger parameter space (the \WC model + nuisances) and the nested one (the \sm, with 0 free parameters + nuisances).  For the 3-parameter fit $n=3$, and for the $C_9$-only test, $n=1$.  Given that our best fit lies far from the edges of the parameter space and the overall sample size is large, assuming the asymptotic limit of Wilks' theorem is a very good approximation under the assumption of normally-distributed errors.}

Previous analysis of older datasets in terms of the same \WCs have assumed that the covariance matrix describing the theoretical uncertainties on the observable predictions could be reliably approximated by its value computed for the \sm, across the entire \WC parameter space. In our fits, we have explicitly recomputed these theoretical uncertainties at every point in the \WC parameter space.  We show the impact of this improvement in \figref{fig::WCresults}, by indicating with a grey star and dashed grey curves the best fit and 1, 2 and $3\sigma$ confidence regions that would result from adopting the \sm approximation.  The central value is not strongly affected, but the impact upon the resulting confidence regions is non-negligible.

\begin{figure}
\hspace{-5mm}
\centering
\includegraphics[width=0.47\textwidth]{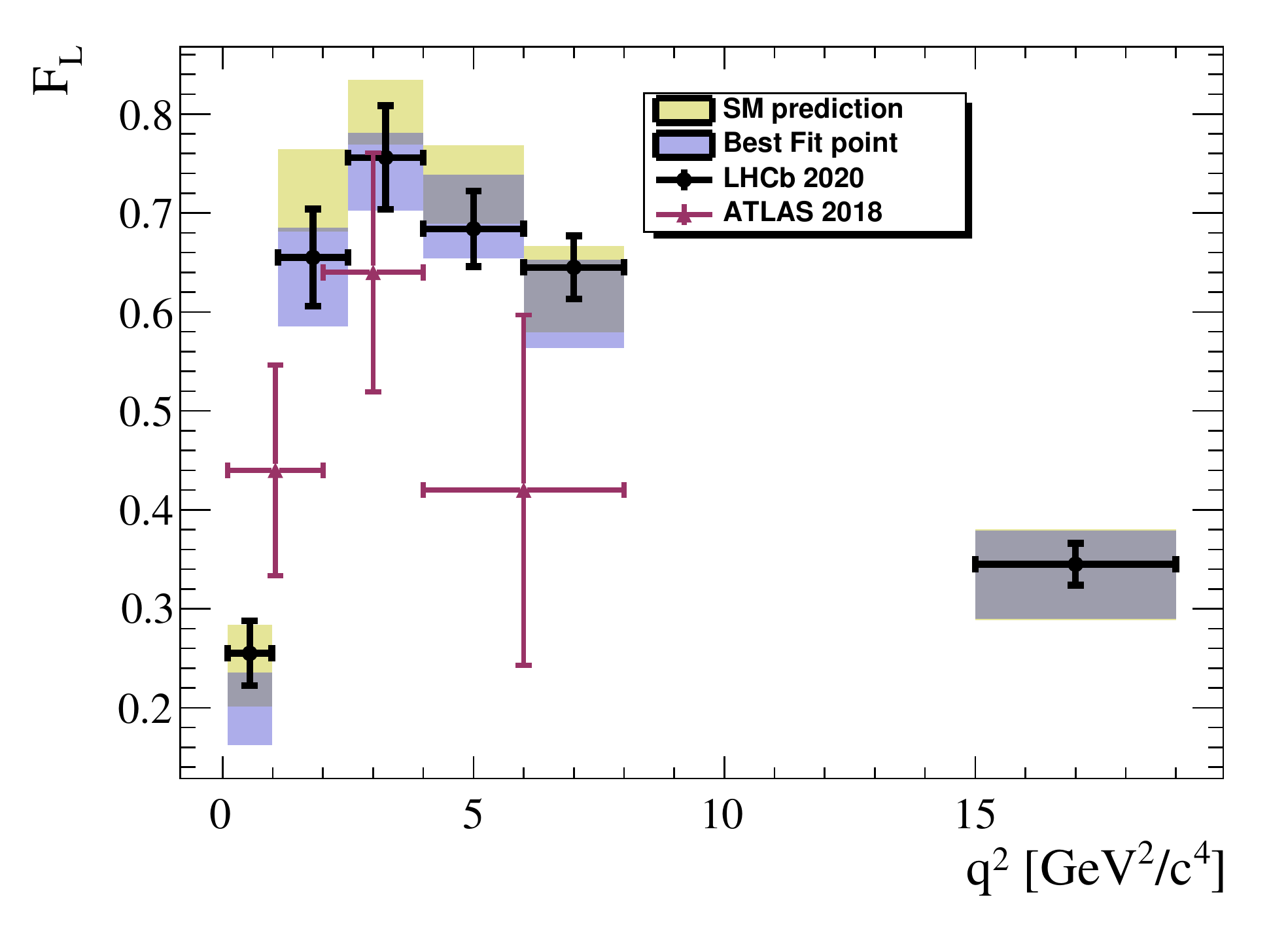}\hspace{5mm}
\includegraphics[width=0.47\textwidth]{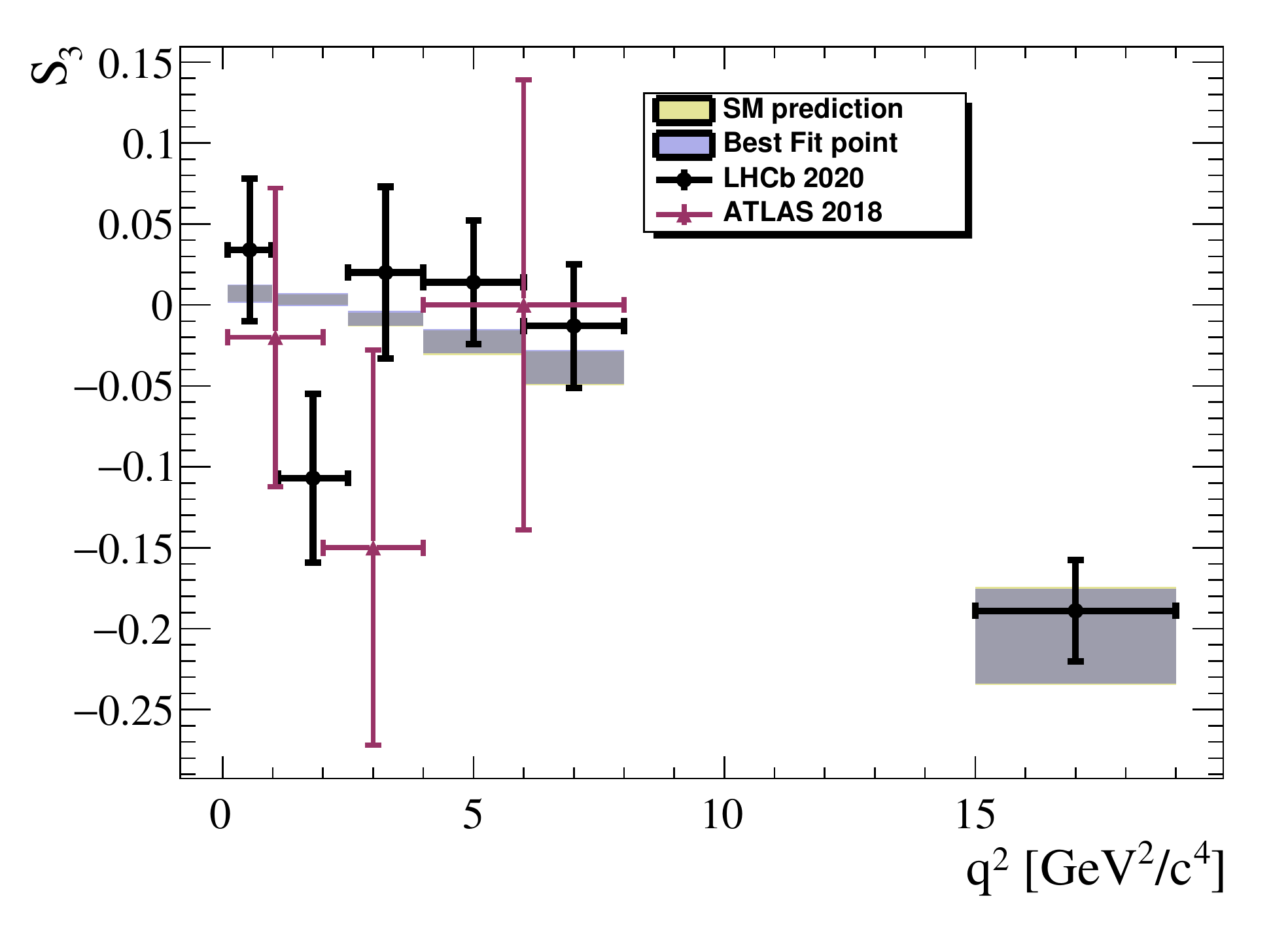}\\
\includegraphics[width=0.47\textwidth]{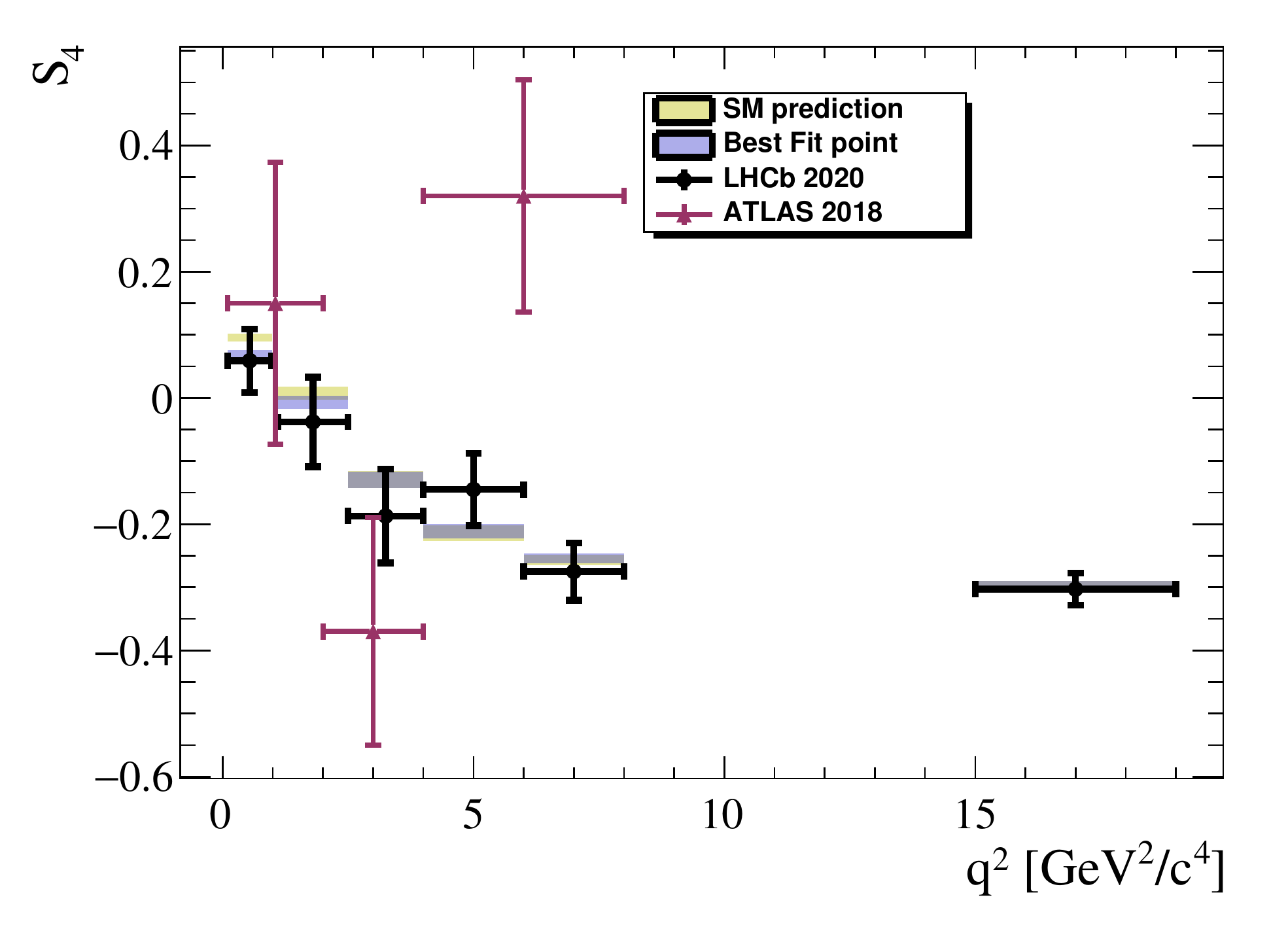}\hspace{5mm}
\includegraphics[width=0.47\textwidth]{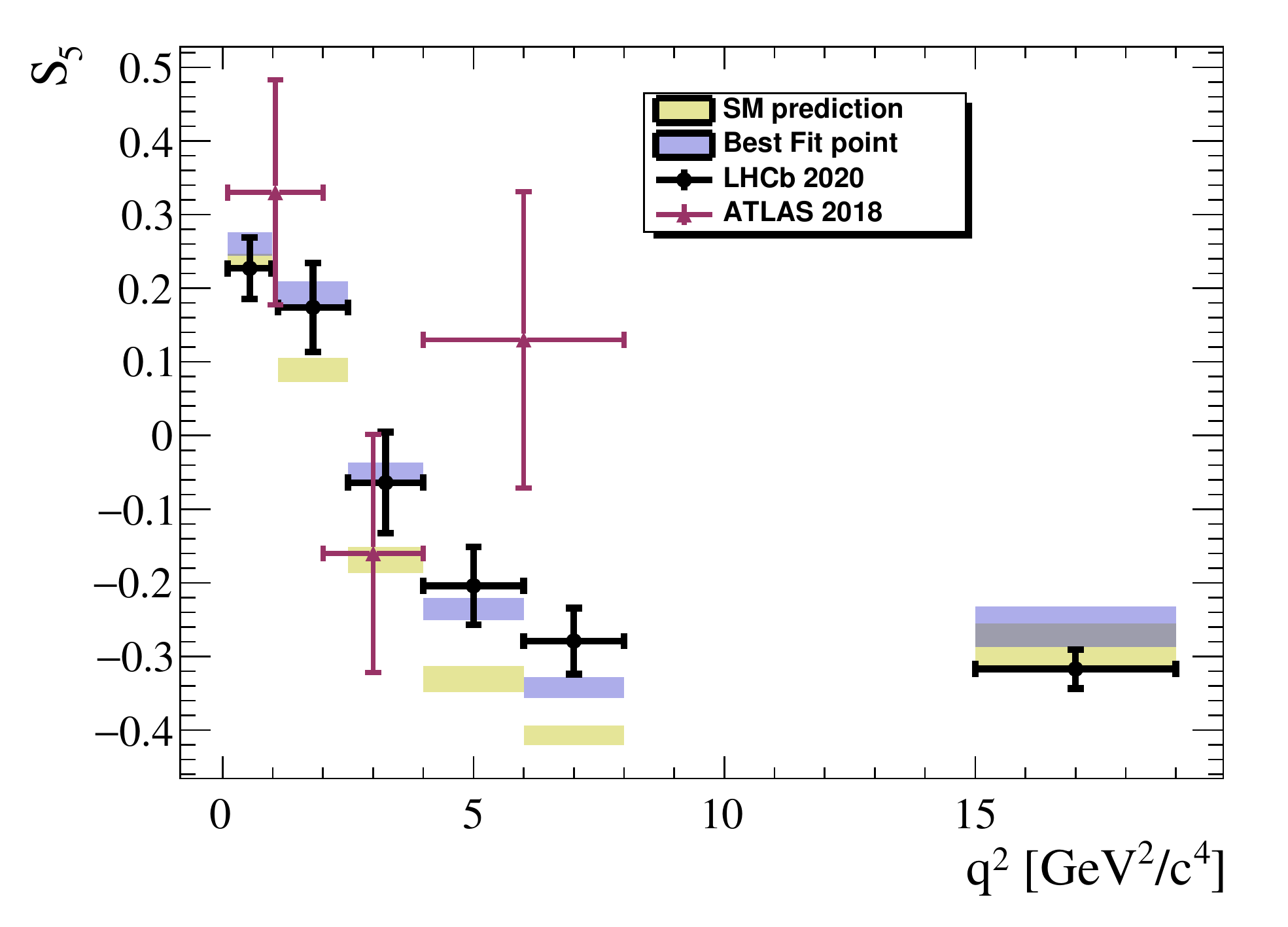}\\
\includegraphics[width=0.47\textwidth]{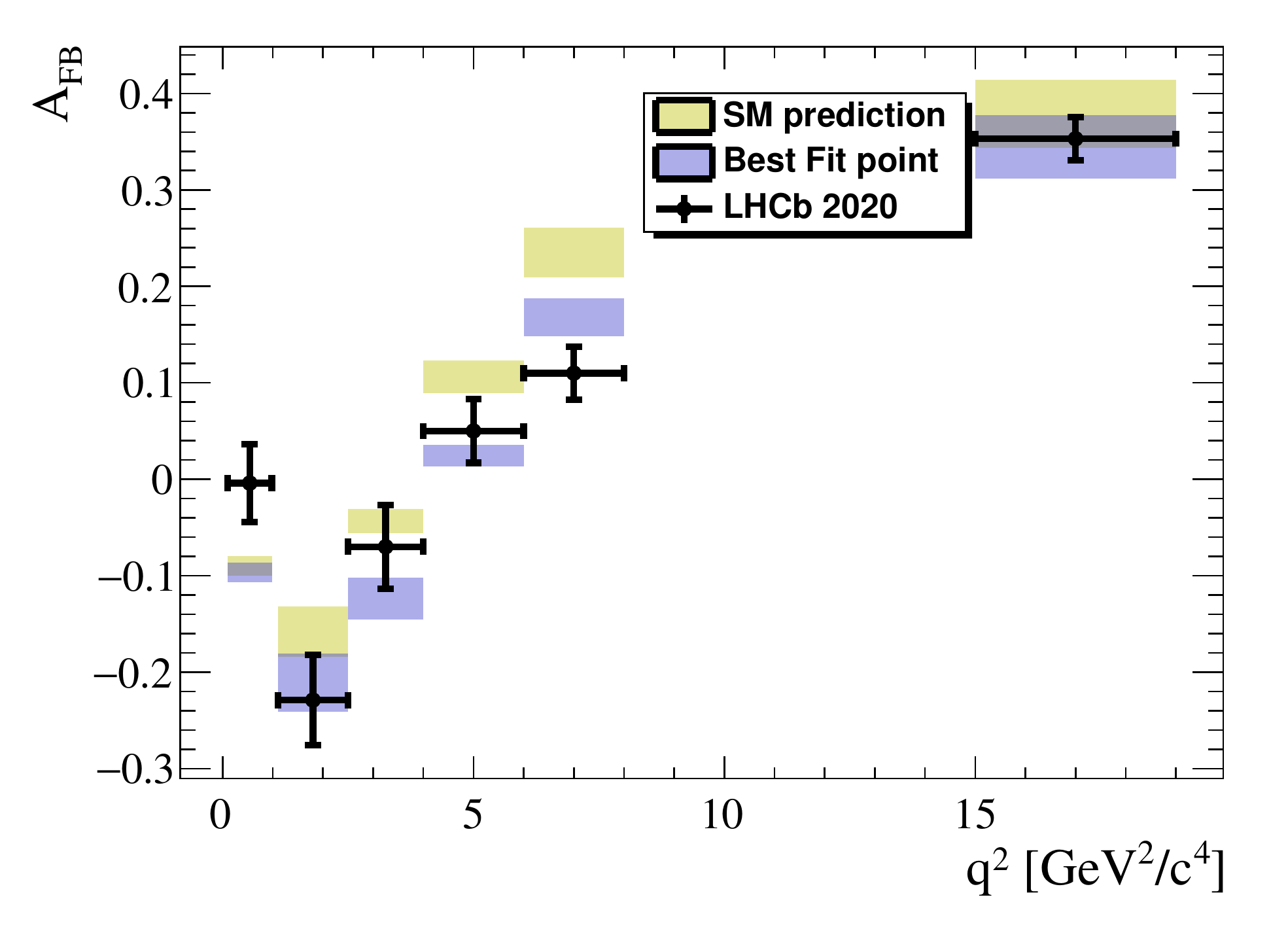}\hspace{5mm}
\includegraphics[width=0.47\textwidth]{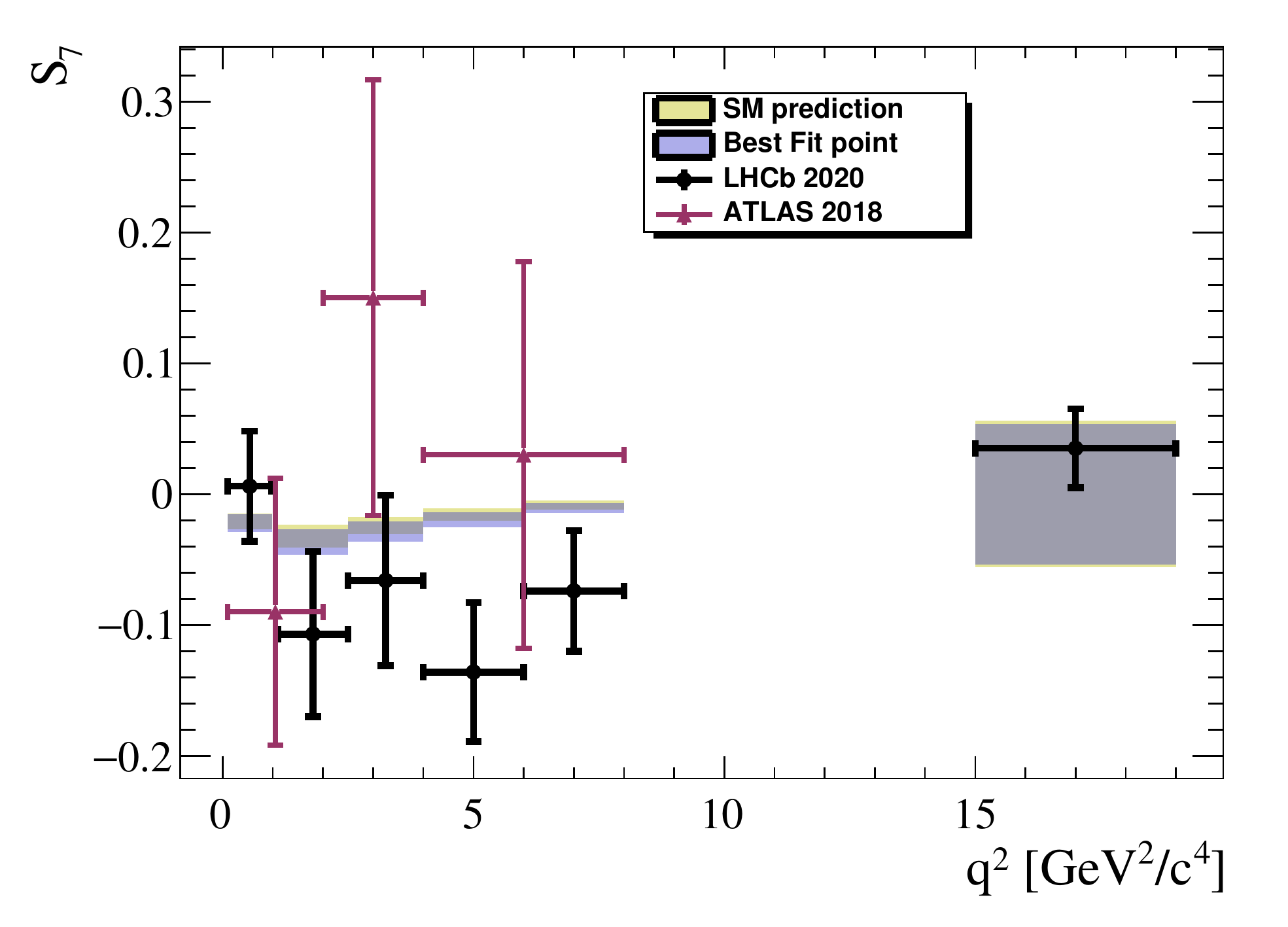}\\
\includegraphics[width=0.47\textwidth]{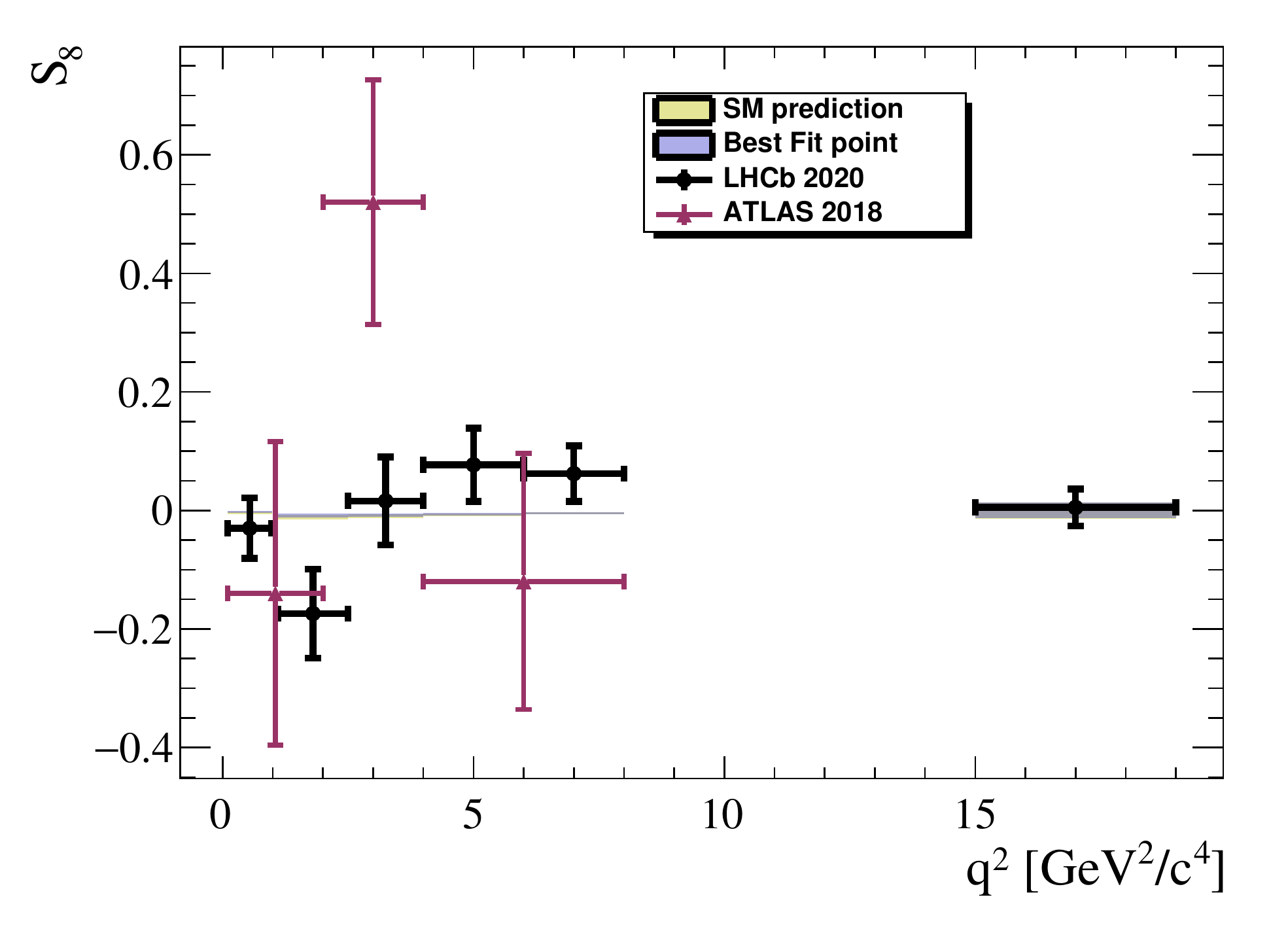}\hspace{5mm}
\includegraphics[width=0.47\textwidth]{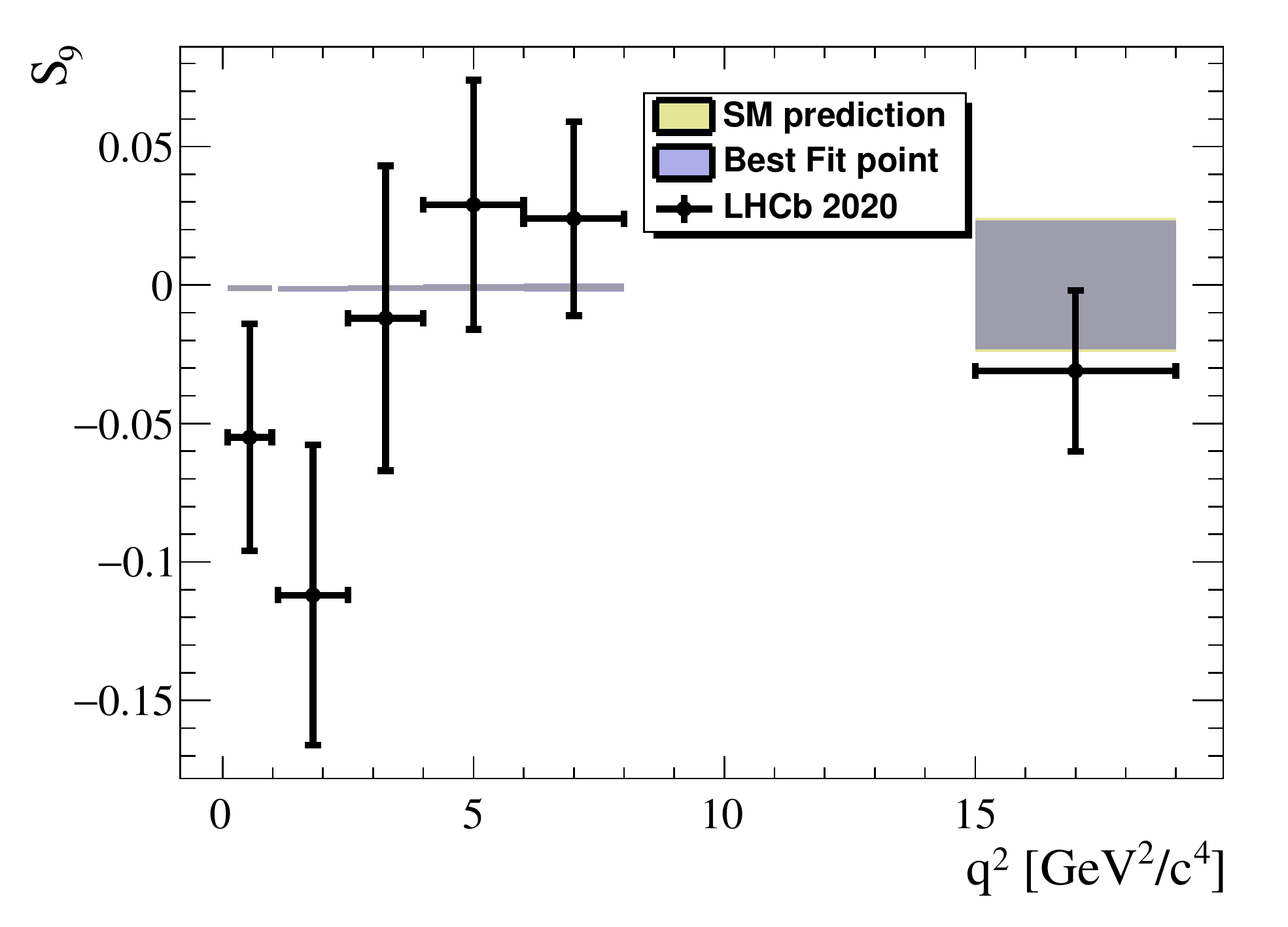}
\caption{Angular observables for the decay \BdToKstarmumu included in our fit. Yellow shading corresponds to Standard Model predictions and uncertainties, and blue shading shows the prediction of our best-fit model. Data points with error bars show measurements from LHCb \cite{Aaij:2020nrf} and ATLAS \cite{Aaboud:2018krd}.  For display purposes, we present theoretical predictions using the same binning as LHCb; in the likelihood function, we always compare experimental results to theory predictions computed in the same binning as used by the experiment in question.
\label{fig::angular}}
\end{figure}

\begin{figure}
\hspace{-6mm}
\centering
\includegraphics[width=0.343\textwidth]{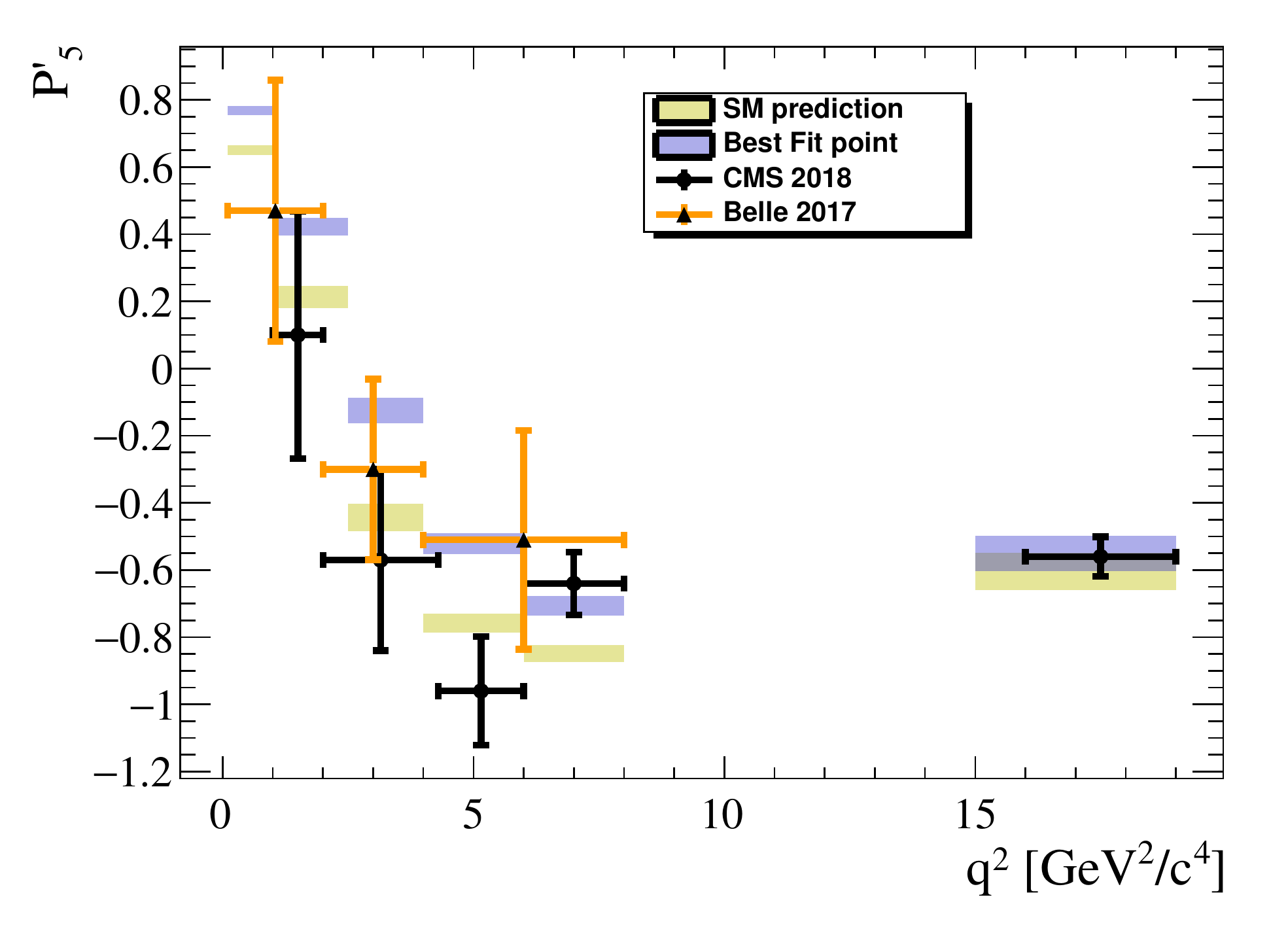}%
\includegraphics[width=0.343\textwidth]{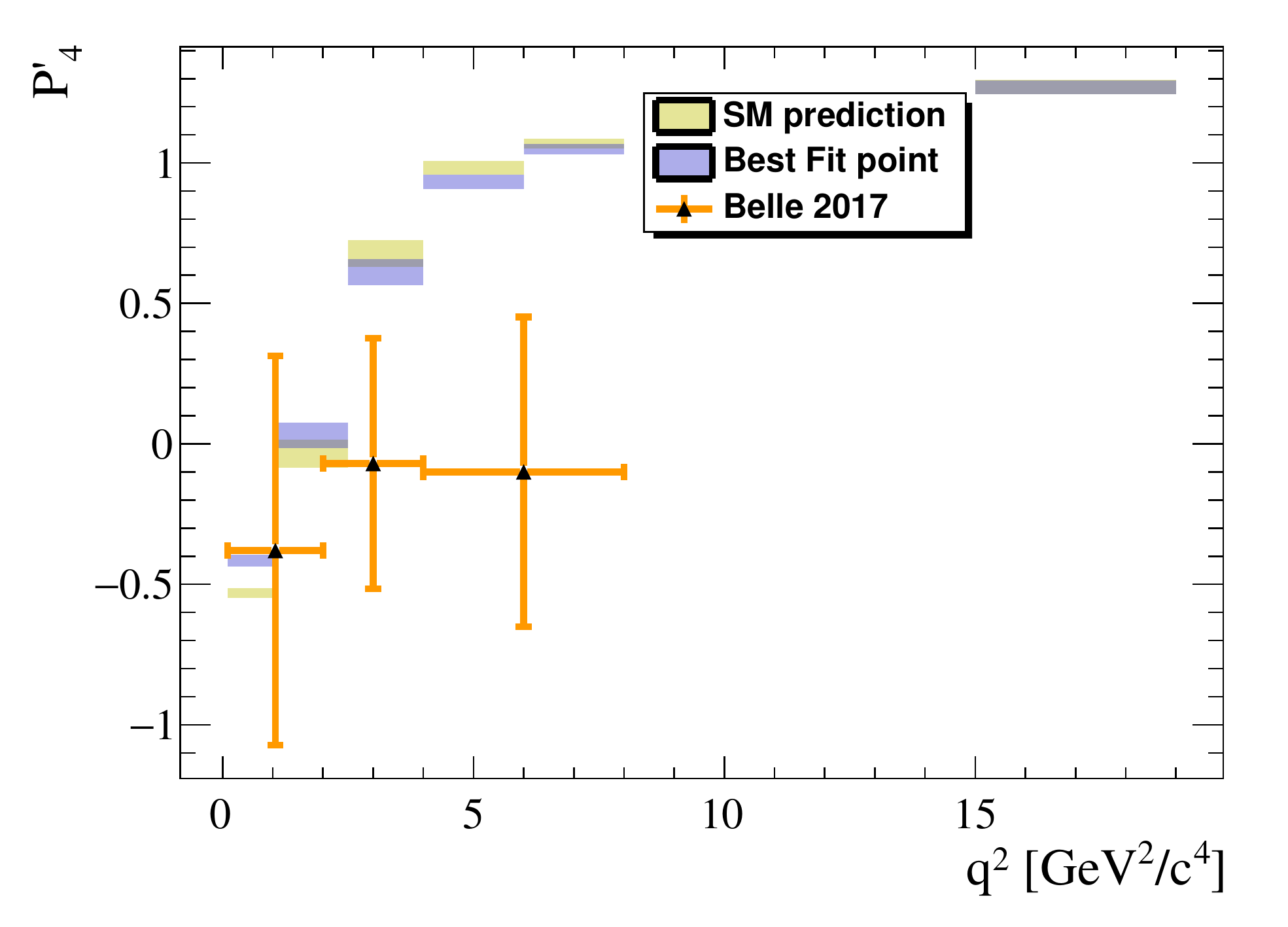}%
\includegraphics[width=0.343\textwidth]{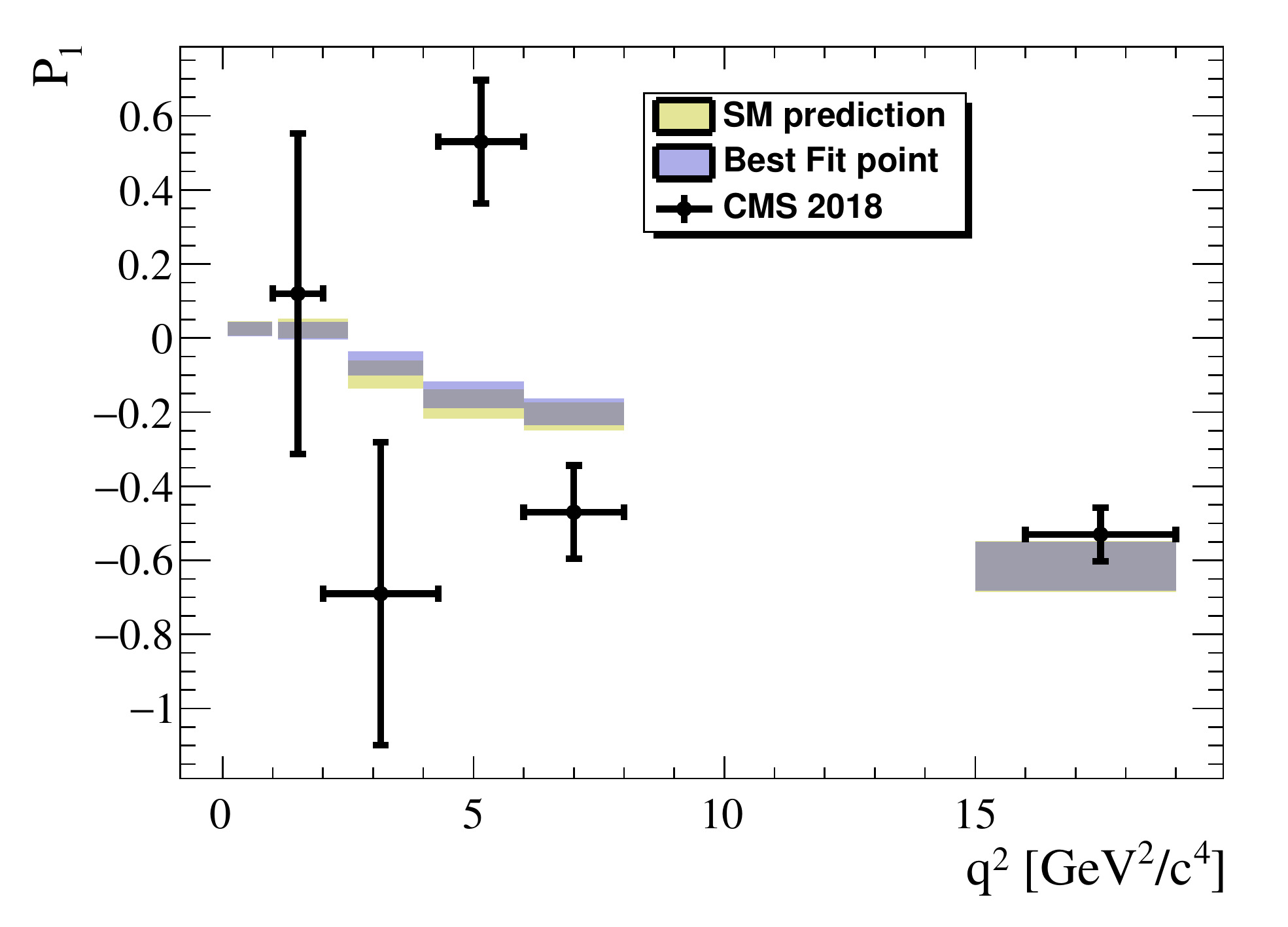}
\caption{Optimised angular observables for the decay \BdToKstarmumu included in our fit. Yellow shading corresponds to Standard Model predictions and uncertainties, and blue shading shows the prediction of our best-fit model. Data points with error bars show measurements from Belle \cite{Wehle:2016yoi} and CMS \cite{CMS:2017ivg}.  For display purposes, we again present the theoretical predictions using the LHCb binning in $q^2$, but consistently recompute the theory predictions in the bins used by Belle and CMS in order to determine their contributions to the overall likelihood.
\label{fig::optimised}}
\end{figure}

\begin{figure}
\centering
\includegraphics[width=0.47\textwidth]{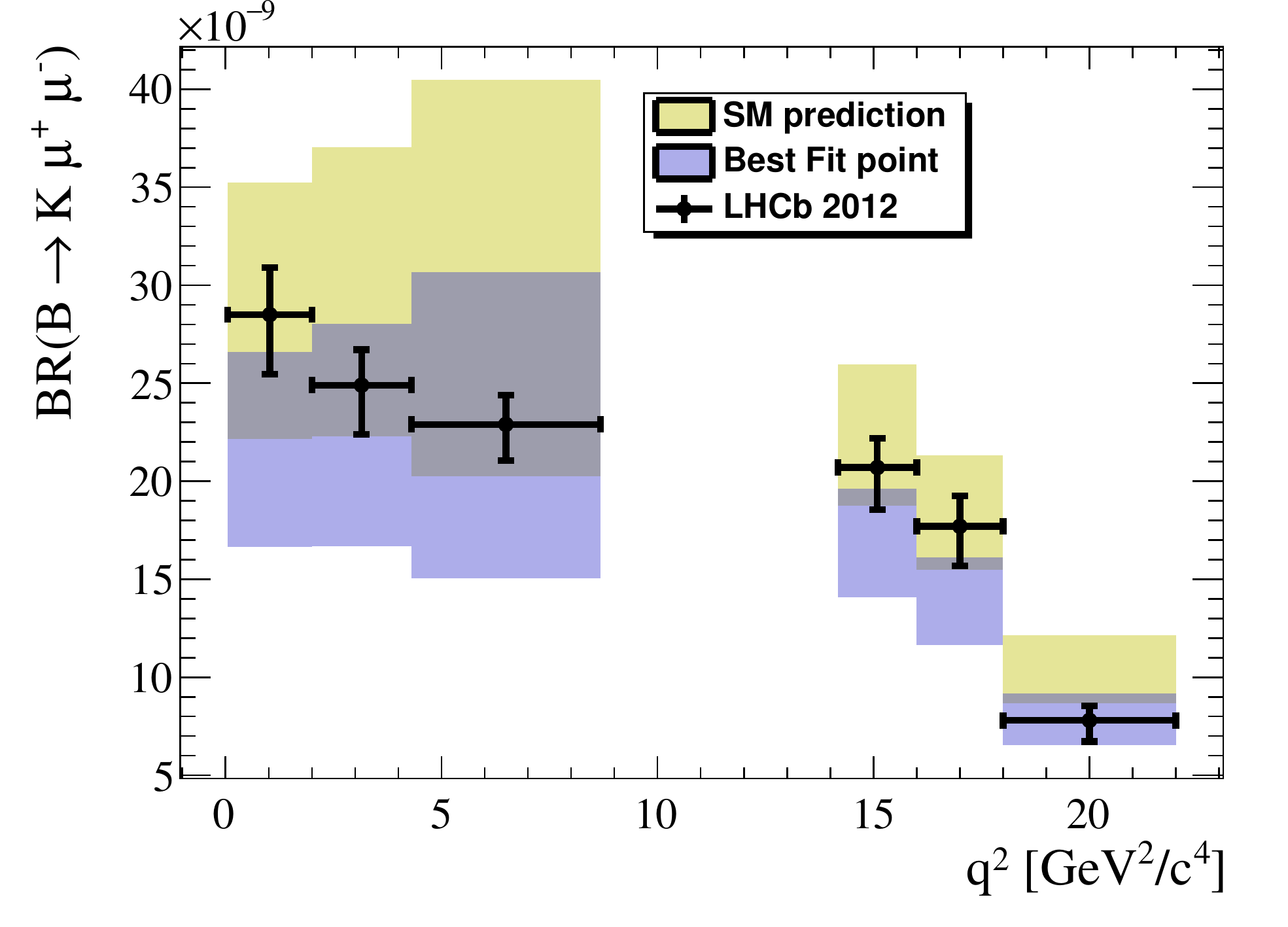}\hspace{5mm}%
\includegraphics[width=0.47\textwidth]{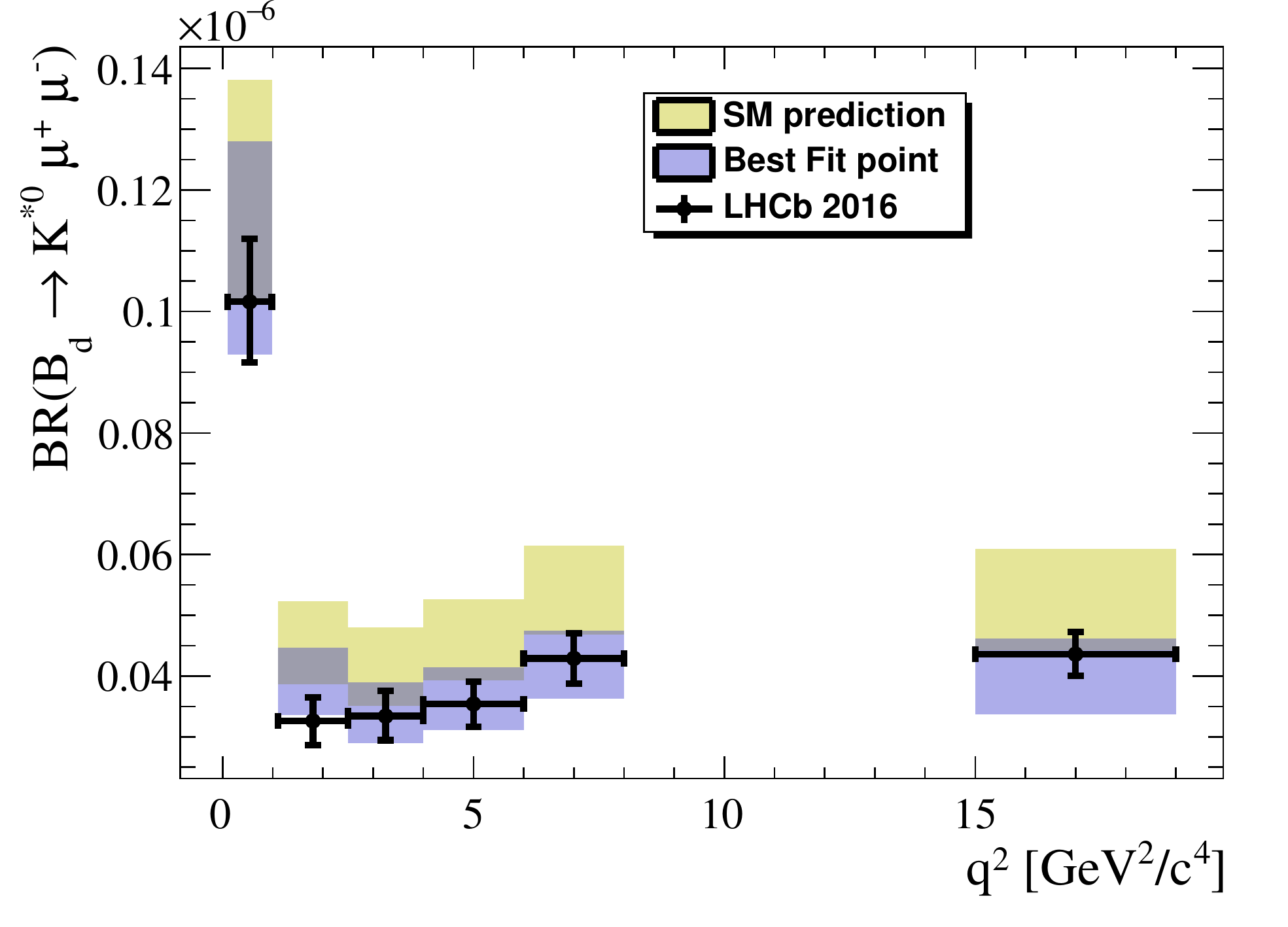}\\
\includegraphics[width=0.47\textwidth]{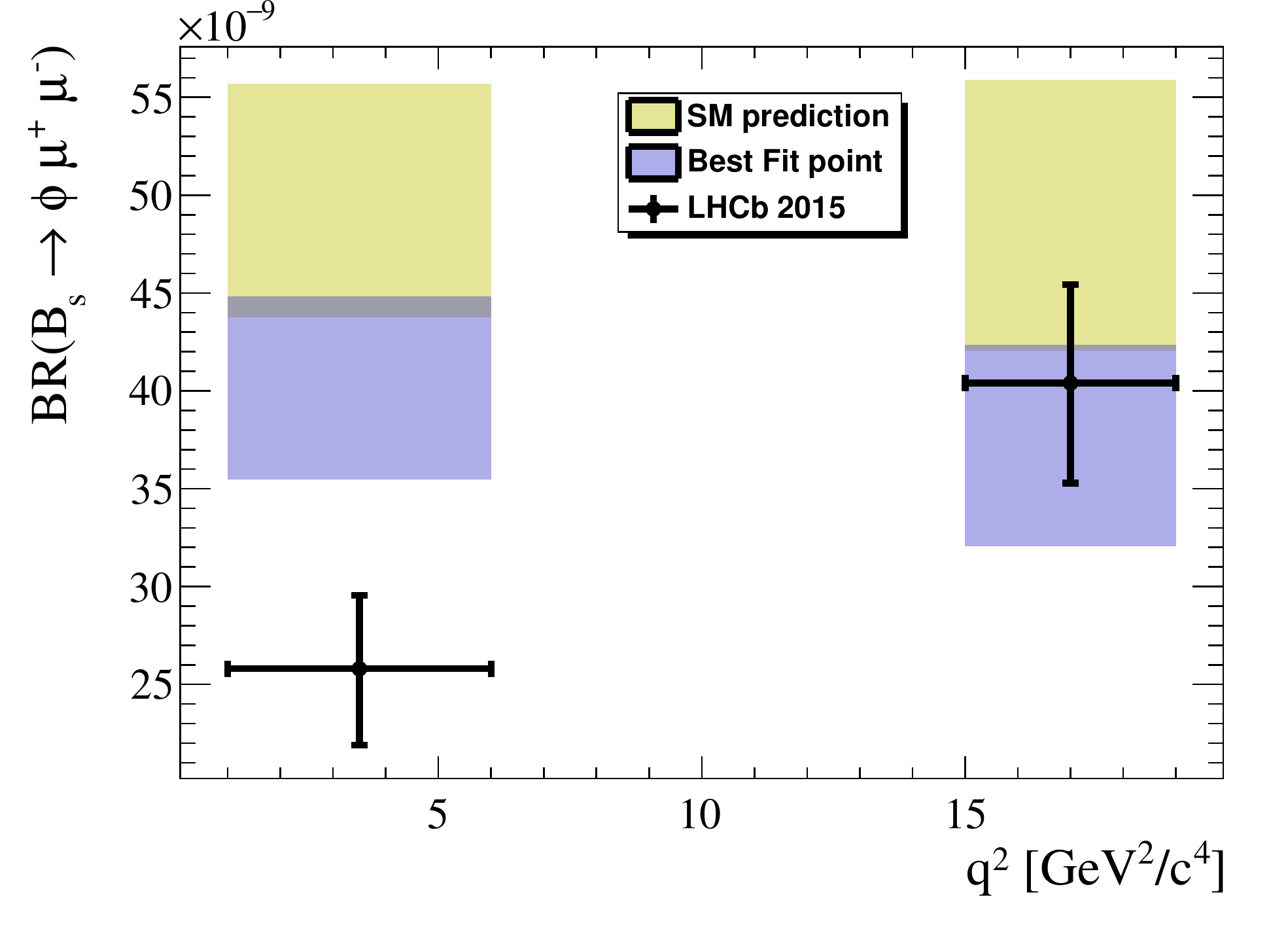}\hspace{5mm}%
\includegraphics[width=0.47\textwidth]{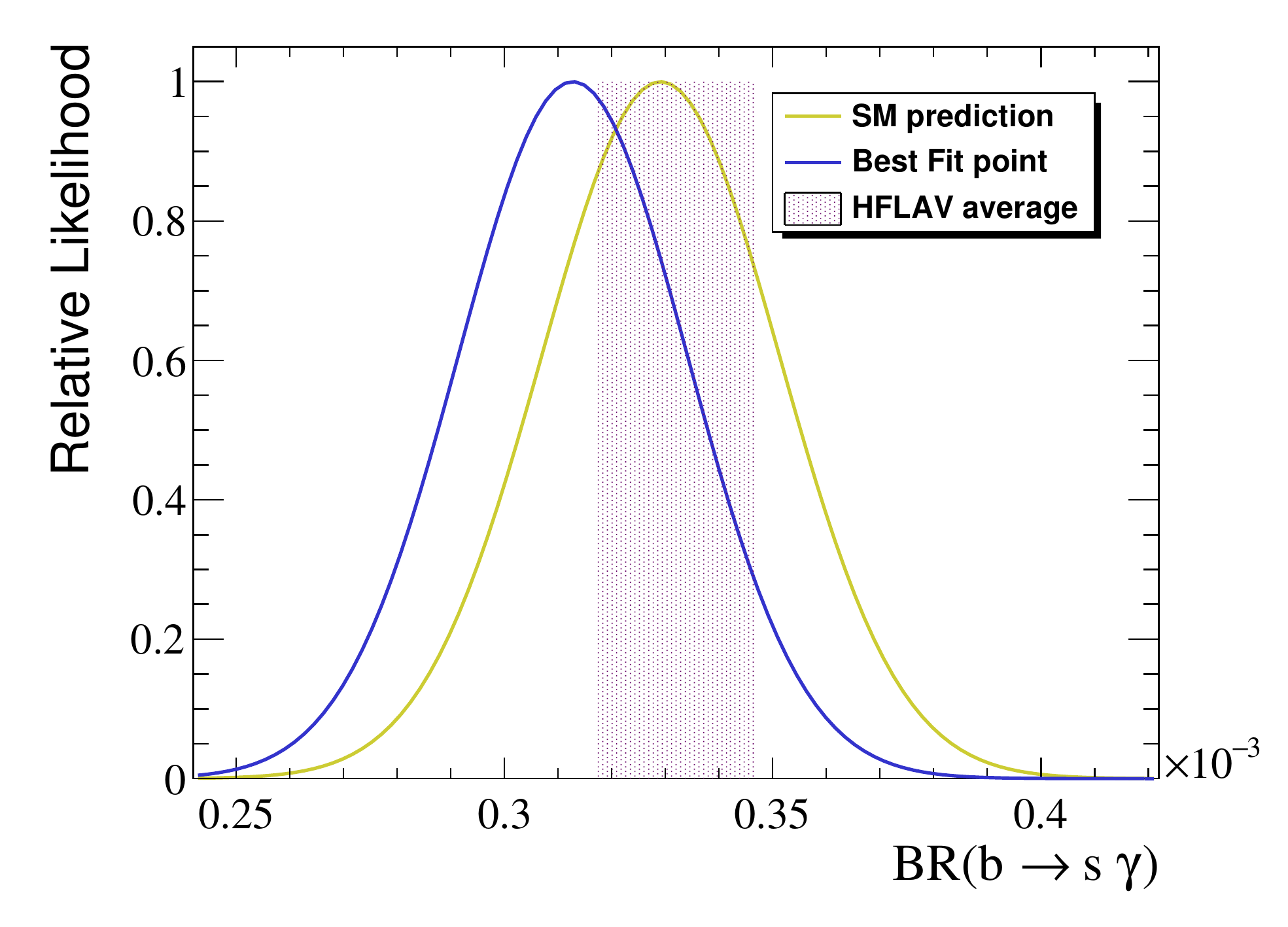}
\includegraphics[width=0.47\textwidth]{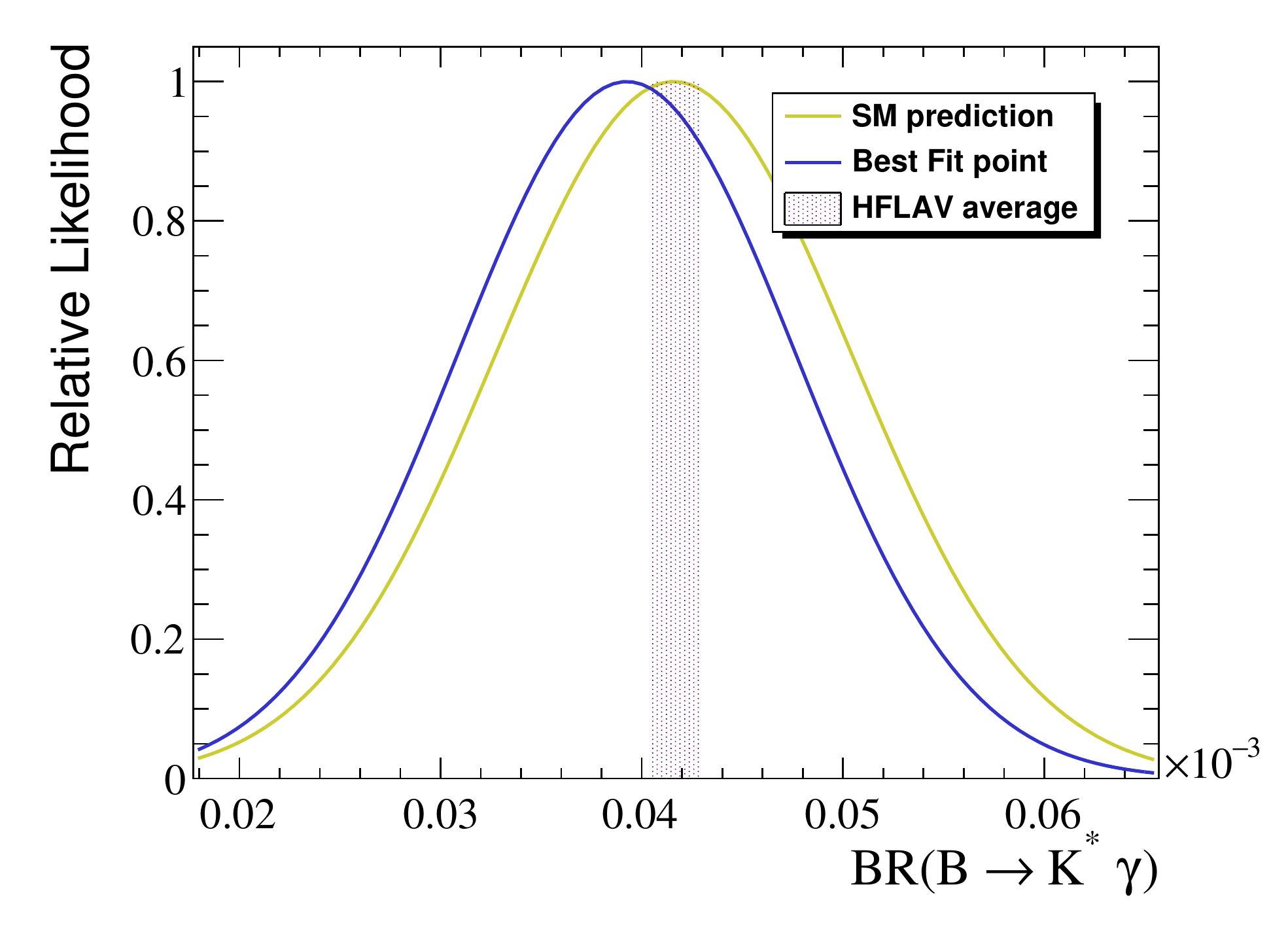}

\caption{Branching fractions for \bTos transitions included in our fit.  Yellow shading corresponds to Standard Model predictions and uncertainties, and blue shading shows the prediction of our best-fit model. Data points with error bars show measurements from LHCb \cite{Aaij:2015esa,Aaij:2016flj,Aaij:2012vr}. The $\bTosgamma$ and $\BtoKstarGamma$ measurements are taken from HFLAV~\cite{Amhis:2018udz}.
\label{fig::br}}
\end{figure}

\begin{figure}
\centering
\includegraphics[width=0.55\textwidth]{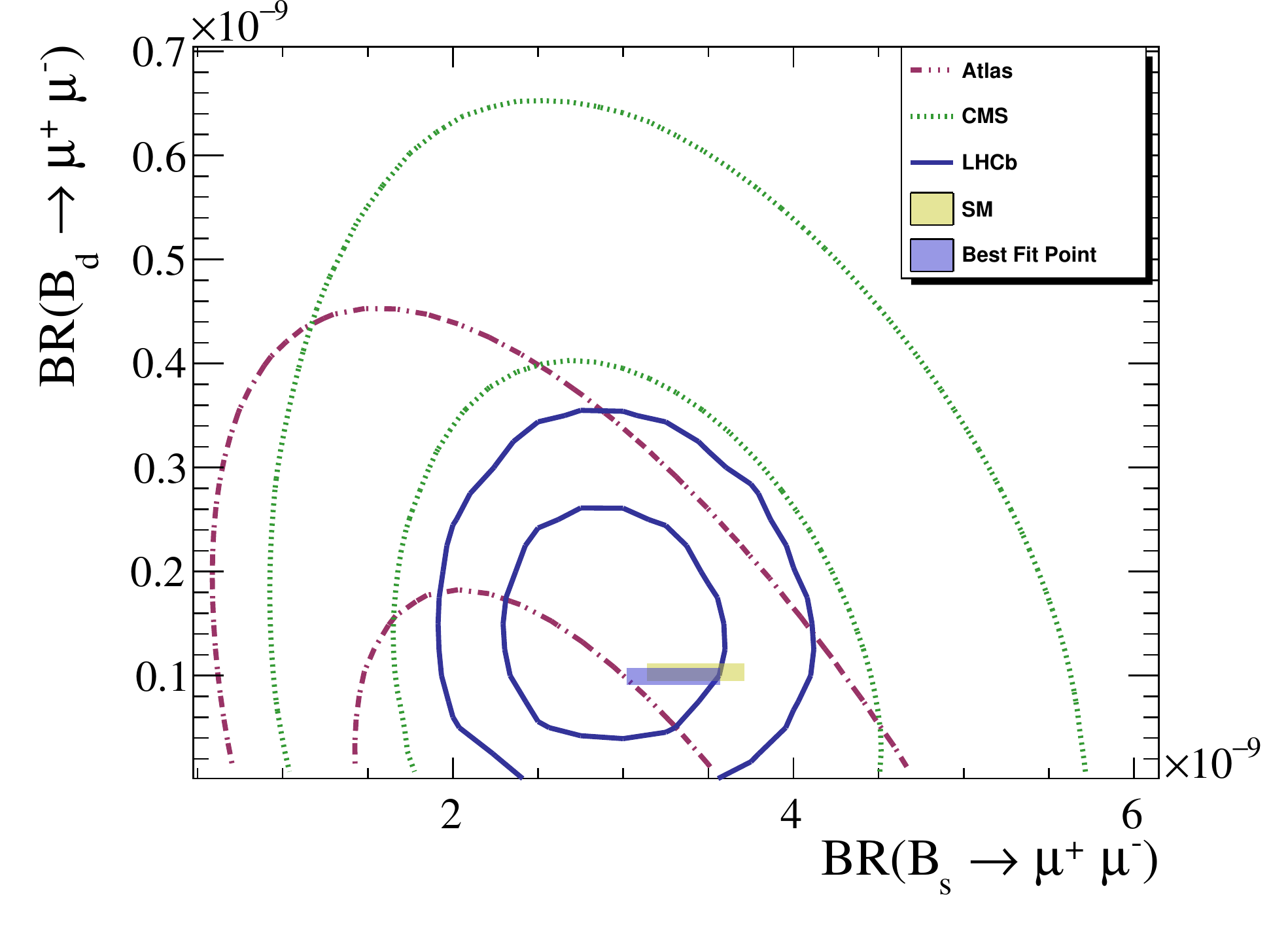}
\caption{Branching fractions for \BTomumu decays included in our fit. Contour lines correspond to the joint 1 and 2$\sigma$ confidence regions obtained in the measurements by ATLAS \cite{Aaboud:2018mst}, CMS \cite{CMS:2019qnb} and LHCb \cite{Aaij:2017vad}.  Yellow shading corresponds to Standard Model predictions and uncertainties, and blue shading shows the prediction and uncertainty of our best-fit model.
\label{fig::B2mumu}}
\end{figure}

In Figs.\ \ref{fig::angular}--\ref{fig::B2mumu}, we provide plots of the key observables and data that enter our fit.  These consist of the \BdToKstarmumu angular observables in the $S_i$ basis (\figref{fig::angular}), their optimised versions (\figref{fig::optimised}), branching fractions for other $b\to s$ processes (\figref{fig::br}), and the joint measurement of the branching fractions of \BsTomumu and \BdTomumu (\figref{fig::B2mumu}).  We show the predictions of both the \sm and our best-fit point, the theoretical uncertainties in each case, and the respective data from LHCb, ATLAS, CMS and Belle used in our fits.  The improvement offered by the best-fit point is most visible in the $S_5$ and $A_{\rm FB}$ angular observables (\figref{fig::angular}), and in the overall branching fractions for \BdToKstarmumu and \BsTophimumu decays (\figref{fig::br}).  Some reduction is expected in the branching fractions for \bTosgamma and \BtoKstarGamma\ relative to the \sm in our best-fit model (\figref{fig::br}), owing to the small positive best-fit value of \recSeven (recalling that the \sm value of $C_7$ is negative).  These are however sufficiently small that the predictions remain consistent with the HFLAV value \cite{Amhis:2018udz}.

\subsection{Implications for future searches}

\begin{figure}
\centering
\includegraphics[width=0.52\textwidth]{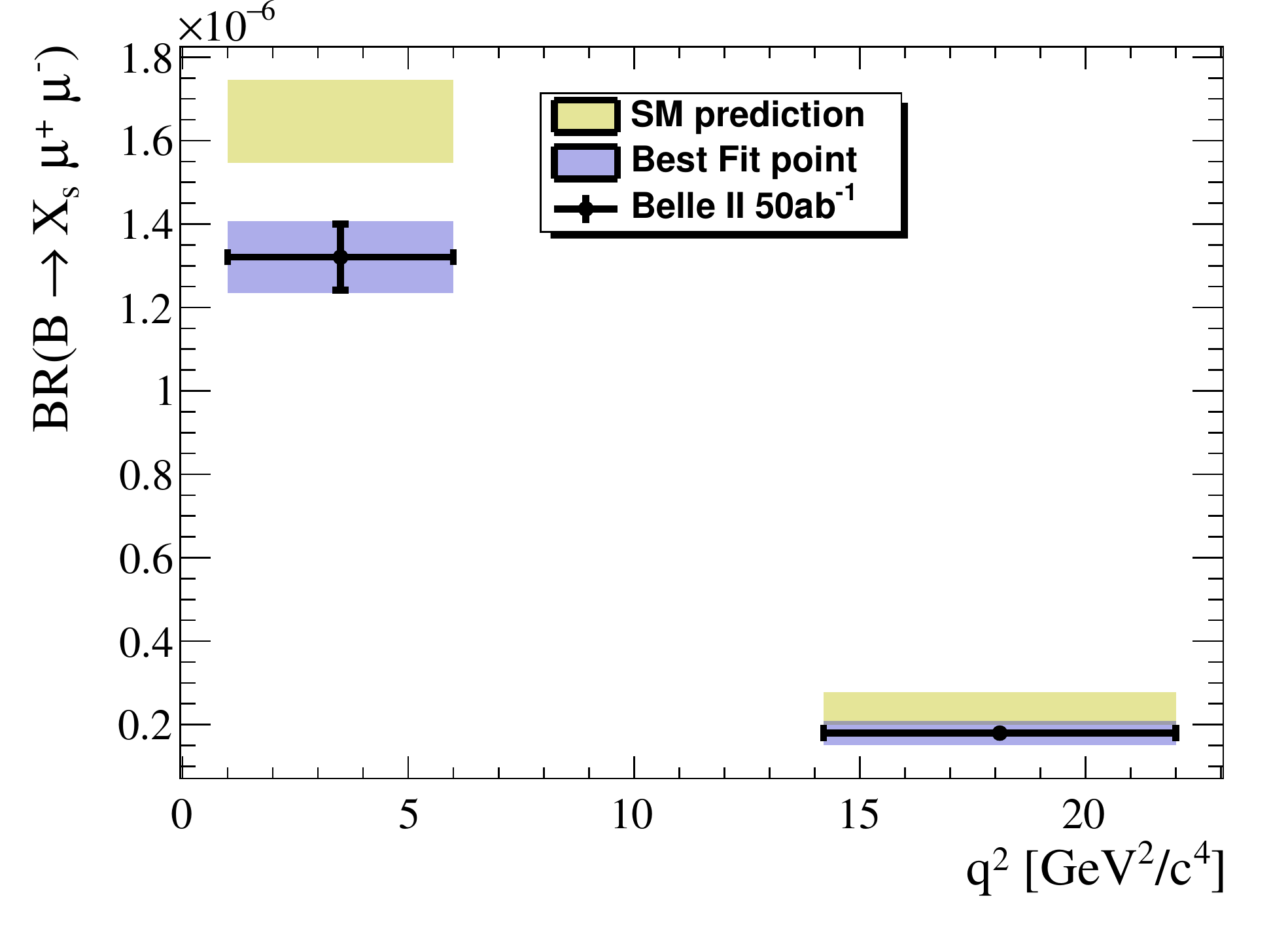}
\caption{Predicted exclusive branching fraction $BR(\BToXsmumu)$ in the Standard model (yellow shading) and in our best-fit model (blue shading).  We also show the expected sensitivity of Belle~II after it has collected 50\,ab$^{-1}$ of data \cite{Kou:2018nap}, assuming that the central measured values are equal to our best-fit prediction.
\label{fig::xsmumu}}
\end{figure}

With higher precision measurements of \BdToKstarmumu in the future, \WC fits will of course  increase in precision. There are however also ongoing efforts to extract the non-factorisable contributions directly from \BdToKstarmumu data \cite{Mauri:2018vbg}. The Belle~II experiment has recently started taking data as well, with the aim of eventually reaching $50~\rm{ab}^{-1}$ integrated luminosity. The unprecedented number of decays that will be contained in this dataset creates the possibility to measure the branching fraction for inclusive \BToXsmumu decays. In \figref{fig::xsmumu}, we show the predictions for $BR(\BToXsmumu)$ in our best-fit model and in the \sm, for two example $q^2$ ranges.  We overlay an expected Belle~II measurement, with the central value set to our best-fit prediction and the uncertainty band based on the predicted sensitivity of Ref.\ \cite{Kou:2018nap}.  Belle~II will clearly have sufficient sensitivity in the low-\qsq region to strongly distinguish the best-fit point from the \sm.

\section{Conclusions}
\label{sec:conclusions}
We have used the \flavbit module of the \gambit package to perform a simultaneous fit of the real parts of the \WCs $C_7$, $C_9$ and $C_{10}$, using combined data on \bTosmumu transitions. Our results show that measurements of flavour anomalies in this sector have reached an intriguing historical juncture, as they are now sufficiently persistent that their tension with the \sm increases steadily as new data are collected. With the inclusion of recently updated results from the LHCb collaboration, we find best-fit values relative to the SM predictions of $\Re(C_7)/\Re(C_7^\text{SM})=0.96$, $\Re(C_9)/\Re(C_9^\text{SM})=0.74$, and $\Re(C_{10})/\Re(C_{10}^\text{SM})=0.99$.  Performing a hypothesis test of the \sm by comparing the log-likelihood of our best-fit point to that of the \sm, we obtain a $6.6\sigma$ preference for our best-fit point over the \sm.  This reduces slightly to $6.3\sigma$ when $C_7$ and $C_{10}$ are profiled out. By explicitly recomputing the theoretical uncertainty covariance matrix at every point in the Wilson coefficient parameter space, we have shown that the best-fit value is not strongly affected by departing from the usual assumption of an SM-only calculation, but that the effect is important for correctly determining confidence intervals, and therefore the overall significance of the result.  Our results still rely on our specific choice of parameterisation of the non-factorisable QCD corrections to many key observables, but the more accurate treatment that we employ of the theory uncertainties across the Wilson coefficient parameter space is an important step forward in improving the test of the \sm hypothesis.

Inspection of the observables that entered our fit indicate that our best-fit point better matches measurements of the $S_5$ and $A_{FB}$ observables than the \sm, in addition to the overall branching fractions for \BdToKstarmumu and \BsTophimumu decays. Other observables are less strongly affected. Localisation of the apparent new physics contribution in a shift of the $C_9$ Wilson coefficient means that the physics should result from a vector coupling of $b$ and $s$ quarks to muons.

\section*{Acknowledgments}

MC and JB are supported by the Polish National Agency for Academic Exchange under the Bekker program, MC by The Foundation for Polish Science (FNP), PS by the Australian Research Council under grant FT190100814 and MJW by the Australian Research Council Discovery Project DP180102209. This research was supported in part by PL-Grid Infrastructure.  We also acknowledge PRACE for awarding us access to Marconi at CINECA, Italy, and Joliot-Curie at CEA, France. MC is grateful for the hospitality of the Institute of Advanced Study, TUM. Fig.\ \ref{fig::WCresults} was produced with \pippi \cite{pippi}.

\clearpage
\bibliographystyle{JHEP_pat}
\bibliography{references}

\end{document}